\documentclass[english,reprint, citeautoscript, aps, prb, nobibnotes, superscriptaddress]{revtex4-1}
\usepackage[T1]{fontenc}
\usepackage[utf8]{inputenc}
\usepackage{verbatim}
\usepackage{amsmath}
\usepackage{amssymb}
\usepackage{amsfonts}
\usepackage{amstext}
\usepackage{graphicx}
\usepackage{float}
\usepackage{siunitx}
\usepackage[hidelinks]{hyperref}
\usepackage{textcomp}
\usepackage{chngcntr}
\usepackage{soul}
\usepackage{color}
\usepackage{upgreek}
\usepackage{tablefootnote}

\definecolor{purple}{rgb}{0.5,0,0.5}

\newcommand{\angstrom}{\textup{\AA}} 
\newcommand{\kcpm}{\frac{\mathrm{kcal}}{\mathrm{mol}}} 
\newcommand{\kcpmpa}{\frac{\mathrm{kcal}}{\mathrm{mol}\cdot \angstrom}} 

\usepackage{multirow}

\usepackage{booktabs}
\usepackage{babel}
\begin{document}

\title{Lightweight and Effective Tensor Sensitivity for Atomistic Neural Networks}

\author{Michael Chigaev}
\affiliation{Theoretical Division, Los Alamos National Laboratory, Los Alamos, NM 87545, USA}
\affiliation{Center for Nonlinear Studies, Los Alamos National Laboratory, Los Alamos, NM 87545, USA}

\author{Justin S. Smith}
\affiliation{Theoretical Division, Los Alamos National Laboratory, Los Alamos, NM 87545, USA}
\affiliation{Center for Nonlinear Studies, Los Alamos National Laboratory, Los Alamos, NM 87545, USA}
\affiliation{NVIDIA, 2788 San Tomas Expy, Santa Clara, CA 95051, USA}

\author{Steven Anaya}

\affiliation{High Performance Computing Division, Los Alamos National Laboratory, Los
Alamos, NM 87545, USA}
\affiliation{Computer, Computational, and Statistical Sciences Division, Los Alamos National Laboratory, Los
Alamos, NM 87545, USA}

\author{Benjamin Nebgen}
\affiliation{Theoretical Division, Los Alamos National Laboratory, Los
Alamos, NM 87545, USA}

\author{Matthew Bettencourt}
\affiliation{NVIDIA, 2788 San Tomas Expy, Santa Clara, CA 95051, USA}

\author{Kipton Barros}

\affiliation{Theoretical Division, Los Alamos National Laboratory, Los
Alamos, NM 87545, USA}
\affiliation{Center for Nonlinear Studies, Los Alamos National Laboratory, Los Alamos, NM 87545, USA}

\author{Nicholas Lubbers}
\email{nlubbers@lanl.gov}
\affiliation{Computer, Computational, and Statistical Sciences Division, Los Alamos National Laboratory, Los
Alamos, NM 87545, USA}

\begin{abstract}
Atomistic machine learning focuses on the creation of models which obey fundamental symmetries of atomistic configurations, such as permutation, translation, and rotation invariances. In many of these schemes, translation and rotation invariance are achieved by building on scalar invariants, e.g., distances between atom pairs. There is growing interest in molecular representations that work internally with higher rank rotational tensors, e.g., vector displacements between atoms, and tensor products thereof. Here we present a framework for extending the Hierarchically Interacting Particle Neural Network (HIP-NN) with Tensor Sensitivity information (HIP-NN-TS) from each local atomic environment. Crucially, the method employs a weight tying strategy that allows direct incorporation of many-body information while adding very few model parameters. We show that HIP-NN-TS is more accurate than HIP-NN, with negligible increase in parameter count, for several datasets and network sizes. As the dataset becomes more complex, tensor sensitivities provide greater improvements to model accuracy. In particular, HIP-NN-TS achieves a record mean absolute error of 0.927 $\kcpm$ for conformational energy variation on the challenging COMP6 benchmark, which includes a broad set of organic molecules. We also compare the computational performance of HIP-NN-TS to HIP-NN and other models in the literature.
\end{abstract}

\maketitle

\section{Introduction}
 Models of potential energy and atomic forces for molecules and materials have wide ranging applications, from drug and materials design to providing scientific insight into a variety of phenomena. Three primary factors drive the research on developing new potential energy functions for obtaining atomic energies and forces: accuracy, computational efficiency, and transferability between chemical systems. Quantum mechanical (QM) methods, which are based on solving the Schr\"odinger equation, tend to be accurate and transferable between systems since these approximate methods are built upon fundamental physical principles. However, QM methods come at a large computational cost, with the most reliable often scaling cubically in system size, or worse, drastically limiting simulation time and length scales. Classical methods are typically linearly scaling, but at the cost of transferability; due to their restrictive physically inspired functional form, classical potentials often need to be tuned or entirely re-fit for each particular application or state-point of interest. Development of accurate, computationally efficient, and transferable model potentials would greatly impact many fields of research within the domains of materials science, drug design, chemistry, and physics.
  
  Machine Learning (ML) methods have proven highly successful for constructing surrogate models across many domains of science and engineering~\cite{Razavi2012, Cai2021, Haghighat2021}. Early models for QM property prediction include the Behler and Parrinello symmetry functions~\cite{Behler2007}, which are used as input into neural networks for property prediction, and the Coulomb matrix representation~\cite{Rupp2012} with kernel ridge regression. Research as far back as the mid-1990s adopted and applied the concept of neural network potentials~\cite{Blank1995NeuralSurfaces,Hobday1999ApplicationsFunctions} for QM property prediction. Recent progress on machine learning methods in chemistry has led to many breakthroughs with surrogate models able to accurately predict myriad QM properties~\cite{Faber2017PredictionError}. One task of particular interest is the construction of potential energy surfaces~\cite{Behler2014RepresentingPotentials}. These models can be used in lieu of force fields for fast molecular dynamics simulations~\cite{Yao2017ThePhysics}, blending the speed of evaluation of trained machine learning models with the accuracy of the underlying training data sets. Recent progress has shown ML model potentials are able to generalize over entire classes of molecules~\cite{Smith2017b, Yao2017ThePhysics, Sifain2018, Nebgen2018, Zubatyuk2019AccurateNetwork}. 
  
  A central challenge in the development of ML potentials is to accurately describe the complete structure of a molecule or material to a machine learning model, while maintaining physics informed constraints such as rotation, translation and permutation symmetries. However, since molecules vary in size and composition, molecular descriptors (descriptors that fully describe all degrees of freedom of an atomic system) are often avoided in favor of descriptors that account for only the local chemical environment of each atom. These approaches~\cite{Smith2017b, Hansen2015MachineSpace, Behler2007, Thompson2015SpectralPotentials,Shapeev2015MomentPotentials,GAPoriginal,SOAPorginal,Kocer19} employ the concept of featurize-then-learn, and can thus be said to be geometrically shallow; the representation of the environment is pre-specified, and a model is learned on top of these descriptor values. Some of these models use explicit many-body evaluation, such as MBTR, BPNN, and ANI~\cite{Huo2017, Behler2015, Smith2017b}. Explicit many-body features invoke computational cost of order $n_\mathrm{neighbors}^{p-1}$, where $p$ is the order of the correlation; pairwise distances correspond to $p=2$, angular terms to $p=3$, and dihedral terms to $p=4$. Such angular and dihedral terms scale rapidly with the number of neighbors $n_\mathrm{neighbors}$ in the chemical environment of an atom. Alternatively, the spectral approach to featurization~\cite{Thompson2015SpectralPotentials,Wood2018ExtendingForm, Shapeev2015MomentPotentials,AtomicClusterExpansion} bypasses the need to evaluate three-body or higher terms directly by constructing invariant quantities out of tensor information from the perspective of atom-centered density. This higher-order information enables shallow learning models that are able to quickly extract many-body information. Nonetheless, the number of higher order descriptors can grow quite quickly when expanded as a complete series~\cite{AtomicClusterExpansion}. Traditionally, geometrically shallow atomic descriptors are fed into machine learning models, e.g. linear or kernel regression models, to predict independent atomic contributions to a property of interest.
  
   Several works have explored \emph{geometrically deep}, end-to-end-learning approaches based on stacked layers~\cite{Schutt2017Quantum-ChemicalNetworks, Schutt2018SchNetMaterials, Unke2019PhysNet:Charges, Lubbers2018, Zubatyuk2019AccurateNetwork, Gilmer2017}. These stacked layers provide information processing that includes an implicit accounting of some long-range information. Most works to date analyze the system purely in terms of the scalar distances between particles~\cite{Schutt2017Quantum-ChemicalNetworks, Schutt2018SchNetMaterials, Unke2019PhysNet:Charges, Lubbers2018, Gilmer2017}. Using only pairwise distances seems limiting relative to the higher order many-body information that could be extracted from an atomic environment.
   Some models, such as  AIMNet~\cite{Zubatyuk2019AccurateNetwork} and DL-MPNN\cite{DLMPNN}, use angular information in combination with geometric depth. 
  
  In this paper we introduce the Hierarchically Interacting Particle Neural Network with Tensor Sensitivities (HIP-NN-TS). This model extends a \emph{geometrically deep} neural network, HIP-NN~\cite{Lubbers2018}, to operate on higher-rank rotational tensors. In this sense, our approach is related to recently proposed architectures such as Tensor Field Networks~\cite{Thomas2018}, Cormorant~\cite{Anderson2019Cormorant:Networks}, and others~\cite{schutt2021equivariant,klicpera2020fast,satorras2021n,tholke2022torchmd,frank2022sokrates}. This class of models introduces the capability to internally use rotational tensors in a way that transforms properly under global rotations. Such models use tensors to capture higher order many-body information from the local environment of each atom, which can lead to improved accuracy. The HIP-NN-TS architecture is a direct and minimal extension of the original HIP-NN model that introduces one new hyperparameter, the maximum tensor rank $\ell_\mathrm{max}$, and using a weight-tying scheme to avoid introducing a large number of new learnable parameters, making it a lightweight addition to the model. When restricted to scalar sensitivities ($\ell_\mathrm{max} = 0$), HIP-NN-TS coincides with HIP-NN.
By selecting $\ell_\mathrm{max} > 0$, we are effectively allowing the model to view each local atomic environment as a truncated multipole (or spherical harmonic) expansion.

 We perform several numerical experiments to explore the benefits of introducing dipolar ($\ell_\mathrm{max} = 1$) and quadrupolar  ($\ell_\mathrm{max} = 2$) sensitivities.
  We show performance benefits for both small and large models on a diverse set of datasets including QM7, QM9, ANI-1x, and ANI1-ccx. Benefits of tensor sensitivities are observable for equilibrium datasets. More striking improvements to accuracy and transferability are observed, however, for models trained to the non-equilibrium ANI-1x dataset. Our model achieves 0.92 $\kcpm$ accuracy for conformational energy variations on the comprehensive COMP6 benchmark set.
  
  To keep the work in the theme of lightweight modeling, we restrict ourselves to studying models which can be trained in a two-day wall time on a single GPU. We also perform tests of the computational throughput of HIP-NN-TS for inference tasks, characterizing the additional expense of tensor sensitivities. We compare HIP-NN-TS with several ML models published in recent literature, and observe a strong combination of computational performance and accuracy.

\section{Methods\label{sec:hipnn}}

\subsection{Brief review of HIP-NN}

We begin by reviewing the original, scalar HIP-NN model and then explain its generalization
to HIP-NN-TS (TS for tensor sensitivity). Here we present only the most salient physical aspects
of HIP-NN, and refer the reader to Ref.~\onlinecite{Lubbers2018} for full details.

The goal is to model the energy $E[\mathbf{r}_{1},\dots\mathbf{r}_{N_{\mathrm{atom}}}]$
of a molecule or a bulk atomic system. It is standard to begin by
decomposing the energy model $\hat{E}$ as a sum over local contributions,
\begin{equation}
E\approx\hat{E}=\sum_{i=1}^{N_{\mathrm{atom}}}\hat{E}_{i}.\label{eq:e_decomp}
\end{equation}
The contribution $\hat{E}_{i}$ is centered on the $i$th atom, and accepts
information about atoms within some fixed interaction range, typically five to ten angstroms. Long-range interactions of fixed form,
such as Coulomb, could be added explicitly. Note that the decomposition
of Eq.~(\ref{eq:e_decomp}) may be non-unique. Nonetheless, such
a decomposition has proven very successful in practice, for both classical
and machine learning-based potentials.

A neural network processes information layer by layer. At each layer of the neural network, HIP-NN produces a characterization (\emph{atomic feature vector}) $z_{i}$ of the local chemical environment at atom $i$. The feature vector has scalar components $z_{i,a}$ for $a=1\dots n_{\mathrm{features}}$, which can be interpreted as neuron activations.  For simplicity, we employ a single hyperparameter $n_{\mathrm{features}}$ yielding the same width for all layers. In the input layer, the species of the atom $i$ is represented in $z_{i,a}$ using a one-hot encoding along dimension $a$.

HIP-NN focuses on the processing of information in \emph{interaction blocks}, which consist of an interaction layer followed by a sequence of $n_\mathrm{layers}$ on-site layers. Each on-site layer consists of a single DNN layer that locally transforms features on an individual atom. Interaction layers allow the atomic feature vectors on neighboring atoms to mix, providing the key step in HIP-NN's operation. An interaction layer transforms its inputs $z$ into activations $z'$ for use by the next layer. The mathematical form is
\begin{equation}
{z'}_{i,a}=f\left(\mathcal{I}_{i,a}(z,\mathbf{r})+\sum_{b}W_{ab}z_{i,b}+B_{a}\right).\label{eq:hipnn-layer}
\end{equation}
The trainable weight matrix $W$ and offset vector $B$
act within a site $i$. The interaction term $\mathcal{I}_{i,a}(z,\mathbf{r})$ provides crucial environmental information. For HIP-NN, the interaction term is
\begin{equation}
\mathcal{I}_{i,a}(z,\mathbf{r}) = \mathcal{I}_{i,a}^{\textrm{HIP-NN}}(z,\mathbf{r})=\sum_{j,b}v{}_{ab}(r_{ij})z_{j,b}.\label{eq:a_hipnn_def}
\end{equation}
This term takes information from the feature vectors $z_{j}$ at neighboring sites $j$ \emph{different} from $i$. This scheme can be viewed as a type of message passing~\cite{Gilmer2017, Schutt2018SchNetMaterials}.
The function $v_{ab}(r_{ij})$ is trainable, and depends on the distance between atoms,
\begin{equation}
r_{ij}=|\mathbf{r}_{i}-\mathbf{r}_{j}|.
\end{equation}
In practice, HIP-NN handles the functional dependence of $v_{ab}$
on $r_{ij}$ by expanding $v_{ab}(r) = \sum_\nu V^\nu_{ab} s^\nu(r)$, where $V^\nu_{ab}$ are trainable model weights, and $s^{\nu}(r)$ are a finite basis of \emph{sensitivity
functions}, which are also trainable~\cite{Lubbers2018}.

Prior works with HIP-NN predicted quantities (e.g., energy or atomic charges) as a linear function of feature vectors $z$ located at the input layer and the end of each interaction block, providing a hierarchical regression scheme, and we follow in suit. Thus, for energy, the full expression for the predicted energy is decomposed over both atoms and interaction blocks as
\begin{equation}
\hat{E}=\sum_{i=1}^{N_{\mathrm{atom}}} \sum_{n=0}^{N_{\mathrm{int}}} \hat{E}^{(n)}_{i}, \label{eq:e_decomp_withhier}
\end{equation}
with hierarchical sub-contribution $\hat{E}^{(n)}_{i}$ for each atom and interaction block in the network. The network can then be regularized to predict an output quantity hierarchically using a function that quantifies the relative magnitude of earlier and later terms of the hierarchical expansion, such as the quantity
\begin{equation}
R=\sum_{i=1}^{N_{\mathrm{atom}}}\sum_{n=1}^{N_{\mathrm{int}}}\frac{(\hat{E}_{i}^{(n)})^{2}}{(\hat{E}_{i}^{(n)})^{2}+(\hat{E}{}_{i}^{(n-1)})^{2}}.\label{eq:hierarchicality}
\end{equation}

\subsection{Generalizing HIP-NN to include tensor sensitivity}

To generalize interaction layers to pass tensor information between atomic sites, we start by rewriting the interaction term, Eq.~(\ref{eq:a_hipnn_def}), in a more suggestive form,

\begin{equation}
\mathcal{I}_{i,a}^{\textrm{HIP-NN}}=\sum_{j}m_{ij,a},\label{eq:a_hipnn}
\end{equation}
where
\begin{equation}
m_{ij,a}=\sum_{b}v_{ab}(r_{ij})z_{j,b}\label{eq:m_def}
\end{equation}
is now interpreted as a scalar message passed from atom $j$ to $i$. At each
interaction layer, the atom $i$ receives information only about pairwise distances $r_{ij}$. In particular, the scalar message $m_{ij,a}$ captures \emph{none }of the angular information carried by the unit vectors 

\begin{equation}
\hat{\mathbf{r}}_{ij}=\frac{\mathbf{r}_{j}-\mathbf{r}_{i}}{r_{ij}}.
\end{equation}

In HIP-NN-TS, messages from $j$ to $i$ are generalized to rank-$\ell$ tensors
\begin{equation}
\mathbf{M}_{ij,a}^{(\ell)}=\mathbf{T}^{(\ell)}(\hat{\mathbf{r}}_{ij})m_{ij,a}.\label{eq:M_def}
\end{equation}
The rank-$\ell$ irreducible Cartesian tensors $\mathbf{T}^{(\ell)}(\mathbf{r})$ can be used to describe functions on the sphere~\cite{Efimov1979}. The first few are,
\begin{align}
T^{(0)}(\mathbf{r}) & =1\label{eq:monopole}\\
T_{\alpha}^{(1)}(\mathbf{r}) & =r_{\alpha}\\
T_{\alpha\beta}^{(2)}(\mathbf{r}) & =r_{\alpha}r_{\beta}-\frac{1}{3} \delta_{\alpha \beta} r^{2} \\
T_{\alpha\beta\gamma}^{(3)}(\mathbf{r}) & =r_{\alpha}r_{\beta}r_{\gamma}-\frac{1}{5}(\delta_{\alpha\beta}r_{\gamma}+\delta_{\alpha\gamma}r_{\beta}+\delta_{\beta\gamma}r_{\alpha})r^{2}\label{eq:octupole}
\end{align}
Here, $\alpha$, $\beta$ and $\gamma$ are Cartesian indices that take on values of $x$, $y$, and $z$ in some coordinate system, $\delta$ is the Kronecker delta (identity matrix), and $r^2$ is the magnitude squared of the input vector $\mathbf{r}$. The irreducible tensors $T^{(\ell)}(\mathbf{r})$  may be familiar as coefficients in the multipole expansion of the Coulomb potential~\cite{Efimov1979}. From a group-theoretic viewpoint, they can be constructed by removing lower-order invariants from the usual rank-$\ell$ tensor product, $\hat{\mathbf{r}}\otimes\dots\otimes\hat{\mathbf{r}}$~\cite{Coope1965}. Note, in particular, that $T^{(\ell)}(\hat{\mathbf{r}})$ is
traceless (contraction of any two indices yields zero) and symmetric, yielding $2\ell+1$ independent components. There is
a close relationship between the irreducible Cartesian tensors $T^{(\ell)}(\hat{\mathbf{r}})$ and the spherical harmonics $Y_{m}^{\ell}(\theta,\phi)$; both can be used as bases for functions on the sphere. The index $\ell$ can be viewed as labeling an irreducible representation of the rotation group, SO(3). The message $\mathbf{M}_{ij,a}^{(\ell)}$ from atom $j$ to $i$ can then be interpreted as passing monopole, dipole, and quadrupole information about feature distributions in the environment for $\ell=0,1,2$, respectively.

In the original HIP-NN scheme, Eq.~(\ref{eq:a_hipnn}), one updates
the atomic features $z_{i,a}$ according to scalar information
from the environment. HIP-NN-TS employs more general \emph{environment
tensors }of rank-$\ell$\emph{,}

\begin{equation}
\boldsymbol{\mathcal{E}}_{i,a}^{(\ell)}=\sum_{j}\mathbf{M}_{ij,a}^{(\ell)}=\sum_{j}\mathbf{T}^{(\ell)}(\hat{\mathbf{r}}_{ij})m_{ij,a},\label{eq:env_tensor}
\end{equation}
with
\begin{equation}
\mathcal{I}_{i,a}^{\textrm{HIP-NN}}= \boldsymbol{\mathcal{E}}^{(0)}_{i,a},
\end{equation}
as a special case.
These tensors characterize the environment at atom $i$ via the accumulation
of messages passed by nearby atoms $j$.

One way to interpret the environment tensors is by introducing an idealized environment {\it function}
\begin{equation}
\mathcal{E}_{i,a}(\hat{\mathbf{r}})=\sum_{j}\delta(\hat{\mathbf{r}}-\hat{\mathbf{r}}_{ij})m_{ij,a},\label{eq:env_func}
\end{equation}
which represents the most general functional dependence on the normalized displacement vector $\hat{\mathbf r}_{ij}$ to each of the neighboring atoms $j$ independently.
Each tensor $\boldsymbol{\mathcal{E}}_{i,a}^{(\ell)}$ captures 
partial information from $\mathcal{E}_{i,a}(\hat{\mathbf{r}})$. In
particular, $\boldsymbol{\mathcal{E}}_{i,a}^{(\ell)}$ can be viewed
as a projection of the function $\mathcal{E}_{i,a}(\hat{\mathbf{r}})$
onto the irreducible tensor $\mathbf{T}^{(\ell)}(\mathbf r)$:
\begin{equation}
    \boldsymbol{\mathcal{E}}_{i,a}^{(\ell)}=\int\mathbf{T}^{(\ell)}(\hat{\mathbf{r}})\mathcal{E}_{i,a}(\hat{\mathbf{r}})\,d^{2}\hat{\mathbf{r}}.
\end{equation}
The complete set of environment tensors ($\boldsymbol{\mathcal{E}}_{i,a}^{(\ell)}$
for $\ell=0\dots\infty$) would faithfully capture all information in the function 
$\mathcal{E}_{i,a}(\hat{\mathbf{r}})$.

It remains to define a HIP-NN-TS generalization of the HIP-NN rule,
Eq.~(\ref{eq:a_hipnn}), for updating atomic features. The high level
goal is to combine information from environment tensors at all ranks
$\ell$. Our approach is to reduce each environment tensor $\boldsymbol{\mathcal{E}}_{i,a}^{(\ell)}$
to a scalar by taking its Frobenius (L2) norm. We define HIP-NN-TS interactions to be a linear combination of contributions at different tensor orders,
\begin{equation}
\mathcal{I}_{i,a}^{\textrm{HIP-NN-TS}}=\boldsymbol{\mathcal{E}}^{(0)}_{i,a}+t^{(1)}_a|\boldsymbol{\mathcal{E}}_{i,a}^{(1)}|+t^{(2)}_a|\boldsymbol{\mathcal{E}}_{i,a}^{(2)}|+\dots\label{eq:hip-vec-1}
\end{equation}

The scalar coefficients $t^{(\ell)}_a$ are trainable parameters. These parameters describe the \emph{mixing} of environmental terms that describe the shape of the atomic environment with respect to different multipole moments indexed by $\ell$.  Any $t_a^{(\ell)}$ could in principle be trained to zero, and consequently, HIP-NN-TS is a strict generalization
of HIP-NN. The coefficient $t_a^{(0)}$ is elided, as it can be rescaled
into the definition of $v_{ab}(r_{ij})$. In the experiments below,
we will truncate the expansion after dipole ($\ell=1$) or quadrupole
($\ell=2$) order.

The HIP-NN-TS architecture is identical to that of HIP-NN except for one important detail: it employs generalized interactions $\mathcal{I}_{i,a} = \mathcal{I}_{i,a}^{\textrm{HIP-NN-TS}}$ in the definition of the interaction layers, Eq.~(\ref{eq:hipnn-layer}).

Importantly, the number of model parameters in HIP-NN-TS is nearly the same as in scalar HIP-NN. In particular, the trainable functions $v_{ab}(r_{ij})$
appearing inside Eq.~(\ref{eq:m_def}) are identical (\emph{weight
tied}) for all ranks $\ell$. The only \emph{new} trainable parameters
in HIP-NN-TS are the scalar mixing coefficients $t_a^{(\ell)}$, which have the same size as the bias parameters $B_a$ for the interaction layers. As such, tensor sensitivities provide significant performance gains using less than 1\% extra parameters in the network, as we will show in our computational experiments.

\subsection{Analysis of tensor sensitivity}

\begin{figure*}
    \centering
    \includegraphics[width=0.8\textwidth]{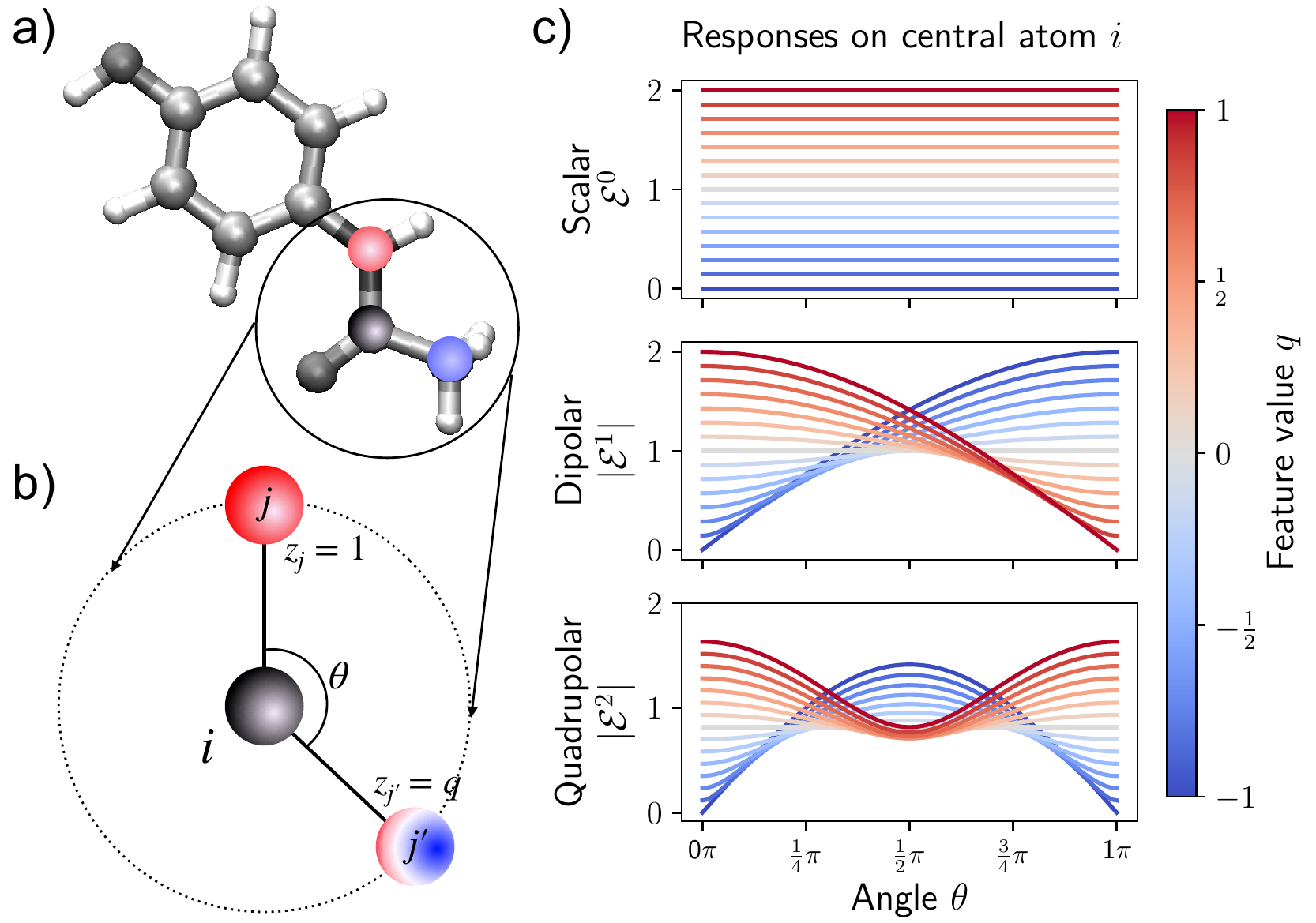}
    \caption{Illustration of how HIP-NN-TS captures angular information about atomic configurations. (a) A molecular geometry containing numerous angles between neighboring atoms, which play a strong role in determining the properties of the molecule. (b) A simple configuration of three atoms---a central atom with two neighbors $j$ and $j'$, described by the angle $\theta$ and the relative feature value $q = z_{j'} / z_j$.  (c) Response of neurons on the central atom $i$ in HIP-NN-TS at different orders of tensor sensitivity $\ell$ as a function of $\theta$ and $q$. The top plot shows the scalar response of the original HIP-NN, which is independent of angle. The bottom two plots show tensor sensitivity information that can be used by HIP-NN-TS to construct richer functions of the atomic environment in its internal representations. The general result is given by Eq.~(\ref{eq:environment_gen}).} 
    \label{fig:response_surface}
\end{figure*}

For the development of intuition, it is interesting to analyze how a single HIP-NN-TS interaction layer is able to extract angular information within a molecule, as depicted in Fig~\ref{fig:response_surface}a. We restrict ourselves to the analysis of a central atom $i$ and messages passed to it via two neighbors $j$ and $j'$, as depicted in Fig~\ref{fig:response_surface}b.  We will focus on a single atom $i$, and a single message feature $a$, which allows to suppress these two indices. Without loss of generality, we assign atom $j$ a feature value of 1, and atom $j'$ a feature value of $q$.  The rank-$\ell$ environment tensors on atom $i$ are
\begin{align}
\boldsymbol{\mathcal{E}}^{(\ell)} & =\mathbf{T}^{(\ell)}(\hat{\mathbf{r}}_{ij})+\mathbf{T}^{(\ell)}(\hat{\mathbf{r}}_{ij'})q.
\end{align}

Our interest is the dependence of $\boldsymbol{\mathcal{E}}^{(\ell)}$
on the angle $\theta$ defined by $\cos\theta=\hat{\mathbf{r}}_{ij}\cdot\hat{\mathbf{r}}_{ij'}$.
Inserting from Eqs.~(\ref{eq:monopole})--(\ref{eq:octupole}),
the first few environment tensors are explicitly,

\begin{eqnarray}
\boldsymbol{\mathcal{E}}^{(0)} & = &1+q\\
\boldsymbol{\mathcal{E}}^{(1)} & = &\hat{\mathbf{r}}_{ij}+q\hat{\mathbf{r}}_{ij'}\\
\boldsymbol{\mathcal{E}}^{(2)} & = &\hat{\mathbf{r}}_{ij}\otimes\hat{\mathbf{r}}_{ij}+q\hat{\mathbf{r}}_{ij'}\otimes\hat{\mathbf{r}}_{ij'}-\frac{1}{3}\left(1+q\right)\mathbf{I}
\end{eqnarray}
where $\mathbf{I}$ denotes the rank-2 identity tensor. Observe that
at the monopole level, the environment scalar $\boldsymbol{\mathcal{E}}^{0}$
discards all angular information as in the original HIP-NN. At the dipole and quadrupole levels, we find
\begin{align}
|\boldsymbol{\mathcal{E}}^{(1)}| & =\sqrt{1+q^{2}+2q\cos\theta_{j,j'}},\label{eq:a1_example}\\
|\boldsymbol{\mathcal{E}}^{(2)}| & =\sqrt{\frac{2}{3}(1-q+q^{2})+2q\cos^{2}\theta_{j,j'}}.\label{eq:a2_example}
\end{align}
The dependence on angle $\theta$ is depicted in Fig.~\ref{fig:response_surface}b for varying $q$ and $\theta$ values. Higher rank $\ell$ yields a sensitivity to higher frequencies of $\theta$. In general, the norm of the rank-$\ell$ environment tensor is
\begin{equation}
|\boldsymbol{\mathcal{E}}^{(\ell)}|=\sqrt{c_{\ell}\left[1+q^{2}+2qP_{\ell}(\cos\theta)\right]},\label{eq:environment_gen}
\end{equation}
where $c_{\ell}=\ell!/(2\ell-1)!!$ and $P_{\ell}(x)$ denotes the Legendre polynomial.
This follows from the tensor contraction identity, $T^{(\ell)}(\mathbf{r})\cdot T^{(\ell)}(\mathbf{r}')=c_{\ell}P_{\ell}(\mathbf{r}\cdot\mathbf{r}')$~\cite{Efimov1979}.

\subsection{Interpretation of tensor sensitivity}

Here we revisit Fig.~\ref{fig:response_surface} under a more qualitative lens. Fig.~\ref{fig:response_surface}a displays a configuration that spans the response of interaction layer including tensor sensitivity: A central particle placed at the origin, with two neighbors at angle $\theta$ with respect to the central atom. The first neighbor has a feature value of 1, and the second neighbor has a feature value of $k$. Fig.~\ref{fig:response_surface}c displays the corresponding response of $\ell=0$ (scalar), $\ell=1$ (dipole), and $\ell=2$ (quadrupole) sensitivity channels. These channels (and potentially, those for even larger $\ell$) can be mixed together to produce a net angular sensitivity within HIP-NN-TS. The angular dependence vanishes for $\ell=0$. The angular dependence for $\ell > 1$ shows maxima and minima, which can be attributed to a kind of constructive or destructive interference between features as a function of $\theta$. Because the tensor sensitivities are calculated via irreducible Cartesian tensors that correspond to spherical harmonics, these functions have similar angular dependence to the standard atomic orbitals surrounding an atom; the $\ell=0,1,2$ sensitivities have a response related to the shapes of $s$, $p$, and $d$ orbitals, respectively. Because chemical bonding can often be explained in terms of combinations of orbitals responding to neighboring orbitals, it is heuristically reasonable to describe the neighborhood of an atom with a low-order expansion in $\ell$. This type of interpretation is not unique to HIP-NN-TS and arguably applies equally well to other tensor-based ML models for atomistic systems.

\section{Results of Computational Studies}

Here we validate and compare the HIP-NN-TS scheme on various existing benchmarks. Each benchmark has a different focus, but all provide an important measure of accuracy for modeling tasks in organic chemistry.

In addition to varying $\ell_\mathrm{max}$, the maximum order of tensor sensitivity in HIP-NN-TS, we vary the three main hyperparameters describing the size and complexity of the network.  The first is $n_\textrm{int}$, the number of interaction blocks in the network. The second is $n_\textrm{layers}$, which gives the number of on-site layers that follow each interaction layer within a block. Last, we vary $n_\textrm{features}$, which controls the layer width, i.e., the number of neurons associated to each atom at each layer in the network. Other details of our training scheme can be found in Appendix Section~\ref{sec:TrainingDetails}.

\textbf{Datasets.} We evaluate the HIP-NN-TS generalization of HIP-NN across several established datasets containing a large variety of organic molecules, including equilibrium and non-equilibrium molecular conformations. We also evaluate HIP-NN-TS on the Allegro-Ag dataset for BCC Silver with a vacancy. A brief summary of the datasets is given in Table~\ref{tab:dset_summary}, and more details are discussed in Appendix Section~\ref{sec:DatasetDetails}.

\begin{table*}[]
    \centering
    \begin{tabular}{c|c|c|c|c|c}
    Name & Size & $N_\textrm{sys}$ & $\langle N_\mathrm{atom}\rangle$ & Elements & Description \\  
    \hline
    QM7\cite{Rupp2012}     & 7K & 7K  & 15 & HCNOS & Relaxed conformations from GDB-7\\
    QM9\cite{Ramakrishnan2014QuantumMolecules}     & 131K & 131K & 18 & HCNOF & Relaxed conformations from GDB-9\\
    ANI-1ccx\cite{Smith2019ApproachingLearning,Smith2020TheMolecules} & 490K & 64K\footnote{Estimated, see appendix~\ref{sec:DatasetDetails}.} & 14  & HCNO & Subsampled from ANI-1x \\
    ANI-1x\cite{Smith2018LessLearning,Smith2020TheMolecules}  & 4.9M & 64K & 15 & HCNO & Active learning with a variety of sources \\
    COMP6\cite{Smith2018LessLearning}   & 101K & 5.6K & 26 & HCNO & 6 test sets designed for use with ANI-1x\\
    QM7-X\cite{qm7x} & 4.2M & 6.9K & 17 & HCNOSCl & Normal Mode Sampling from GDB-7 \\
    Allegro-Ag\cite{allegro}   & 1K & 1 & 71 & Ag & \textit{ab initio} molecular dynamics trajectory \\

    \end{tabular}
    \caption{Summary of datasets considered in this work. `Size' gives the number of configurations, $N_\textrm{sys}$ the number of chemical systems, $\langle N_\textrm{atoms}\rangle$ the average number of atoms per frame, `Elements' the set of atomic elements in the dataset. }
    \label{tab:dset_summary}
\end{table*}

\subsection{QM7, QM9, and ANI1-ccx}
\label{sec:qm7qm9ani1ccx}

\textbf{Model searches on QM7, QM9, and ANI-1ccx.} Table~\ref{tab:dataset_overview} shows the results of hyperparameter searches for the three benchmark datasets, with the restriction to $\ell_\mathrm{max} \in \{0 , 1\}$. The optimal models were selected according to mean absolute error (MAE) performance. Both MAE and root mean square error (RMSE) are reported. Each model architecture was studied with 8 training/testing splits of the data, and results on the test data were averaged across these trials. We performed exhaustive grid search in the space $n_{\mathrm{features}} \in \{10,20,30,40,60,120\}$, $n_\mathrm{int} \in \{1,2,3\}$, and $n_\mathrm{layers} \in \{2,3,4\}$. Due to the larger size of the ANI-1ccx database, we limited the search to $n_{\mathrm{int}} \in \{1,2\}$ and $n_{\mathrm{features}} \in \{20,60,120\}$. In all cases, HIP-NN-TS with $\ell_\mathrm{max}=1$ outperformed the original HIP-NN architecture, i.e., $\ell_\mathrm{max}=0$. Tensor sensitivity was most beneficial on the ANI-1ccx benchmark of non-equilibrium molecular energies, and an MAE reduction of approximately 25\% was observed. In all cases, 120 features (the maximum number tested) performed the best. This was unexpected for the relatively small QM7 dataset, since more parameters yield a larger model capacity, presumably increasing the danger of overfitting. 
The QM7 tests did, however, benefit from having only a single interaction block, $n_\mathrm{int}$; increasing the number of interaction blocks degraded performance on this small dataset.

\begin{table*}
    \centering
\begin{tabular}{l|l|l|l|lll}
\toprule
         Dataset & $\ell_\mathrm{max}$  &                    MAE ($\kcpm$)    &                   RMSE ($\kcpm$) & $n_\mathrm{int}$ & $n_\mathrm{features}$ & $n_\mathrm{layers}$ \\
\midrule
\multirow{2}{*}{QM7} & 0 &  1.364 $\pm$ 0.026 &  2.312 $\pm$ 0.055 &                  1 &      120 &             2 \\
         & 1 &  1.207 $\pm$ 0.011 &  2.061 $\pm$ 0.023 &                  1 &      120 &             2 \\
\cline{1-7}
\multirow{2}{*}{QM9} & 0 &  0.236 $\pm$ 0.001 &  0.548 $\pm$ 0.007 &                  2 &      120 &             2 \\
      & 1 &  0.205 $\pm$ 0.002 &  0.458 $\pm$ 0.003 &                  2 &      120 &             3 \\
\cline{1-7}
\multirow{2}{*}{ANI-1ccx} & 0 &  2.827 $\pm$ 0.003 &  4.074 $\pm$ 0.005 &                  2 &      120 &             4 \\
      & 1 &  2.116 $\pm$ 0.005 &  3.121 $\pm$ 0.007 &                  2 &      120 &             4 \\
\bottomrule
\end{tabular}

    \caption{Performance of HIP-NN-TS on QM7, QM9, and ANI1-ccx datasets for hyperparameters $n_\mathrm{int}$, $n_\mathrm{features}$ and $n_\mathrm{layers}$ selected by grid search. HIP-NN-TS with $\ell_\mathrm{max} = 1$ consistently outperforms scalar HIP-NN ($\ell_\mathrm{max} = 0$) in both mean absolute error (MAE) and root mean squared error (RMSE). The improvement is more dramatic on the larger ANI-1ccx dataset, which contains non-equilibrium conformers rather than minimum energy conformers only.}
    \label{tab:dataset_overview}
\end{table*}

In our tests, we found no benefits to introducing a third interaction layer ($n_\mathrm{int}=3$), demonstrating that deeper models are not always better. This is consistent with literature that shows that wider networks are preferable to deeper ones for certain problems~\cite{Lee2017, Novak2020, Neyshabur2014, Novak2018, Neyshabur2019}. Nonetheless, the optimal number $n_\mathrm{layers}$ of on-site layers grew with the size of the dataset, showing that a deeper network can be beneficial when sufficient data is available.

\begin{table}
    \centering
\begin{tabular}{lccc}
\toprule
 & Geometric & Sensitivity &  \\
Method & Depth & Order & MAE \\

\midrule

Allegro~\cite{allegro} & 3  & 2 & 0.11 \\

PaiNN~\cite{schutt2021equivariant} & 3  & 1 & 0.13 \\
ET~\cite{tholke2022torchmd} & 8 & 1 & 0.14 \\
DimeNet$^{++}$~\cite{klicpera2020fast} & 6 & 5 & 0.15  \\
DL-MPNN\cite{DLMPNN} & 7 & - & 0.17 \\
PhysNet~\cite{Unke2019PhysNet:Charges} & 5 & 0 &  0.19 \\
{\textbf{HIP-NN-TS}} & 2 & 1 & 0.205 \\
{HIP-NN}  & 2 & 0 & 0.236 \\
{EGNN}~\cite{satorras2021n}  & 7 & 0 & 0.25 \\

GM-sNN~\cite{Zaverkin2020} & 1 & 3 & 0.27 \\
{ANI} & 1 & - &  0.29 \\
SchNet~\cite{Schutt2018SchNetMaterials} & 6 & 0 & 0.31 \\
{\textbf{HIP-NN-TS}}  & 1 & 1 &  0.33 \\
{HIP-NN} & 1 & 0  &  0.39 \\
MPNN~\cite{Gilmer2017} & 3+ & 0 & 0.42 \\
Cormorant~\cite{Anderson2019Cormorant:Networks} & 4 & 3 &  0.51 \\
\bottomrule
\end{tabular}
    \caption{Performance of HIP-NN-TS with $\ell_\mathrm{max}=1$ compared to other published neural networks, as measured by MAE performance on atomization energy predictions for the QM9 dataset, in units of $\kcpm$. `Geometric depth' (analogous to $n_\mathrm{int}$) measures the number of times activations from the atomic environment are propagated between atoms. `Sensitivity order' (analogous to $\ell_\mathrm{max}$) measures the tensor order at which a layer processes information.    ANI and DL-MPNN are difficult to classify this way, see the text for remarks. Models without a citation given were trained in this work.}
    \label{tab:qm9_comparison}
\end{table}

 \textbf{QM9 in context with the literature.} Table~\ref{tab:qm9_comparison} compares HIP-NN-TS models with several other neural networks on the QM9 benchmark. We categorize the models in terms of a \emph{geometric depth}, the number of times that the network processes the atomic environment during energy prediction. In HIP-NN and HIP-NN-TS this is the number of interaction layers, $n_\mathrm{int}$. We also categorize models in terms of the tensor \emph{sensitivity order}, which is analogous to the $\ell_\mathrm{max}$ of HIP-NN-TS. Loosely speaking, this corresponds to a spherical harmonic order. In the case of the ANI and DL-MPNN architectures, explicit angular functions capture information from the three-body angle distribution, which cannot be simply characterized by a tensor order.
 
 With a geometric depth of two interaction blocks, scalar HIP-NN already performs well, and HIP-NN-TS shows a marked improvement. The performance of these models remains reasonable even with a single interaction layer, which significantly reduces the number of model parameters and effective cutoff radius. Interestingly, we do not observe a consistent trend between performance and hyperparameters such as geometric depth or sensitivity order. This perhaps motivates future work to better understand the performance characteristics of different model architectures as a function of depth and sensitivity order.

\begin{figure*}
    \centering
    \includegraphics[width=0.7\textwidth]{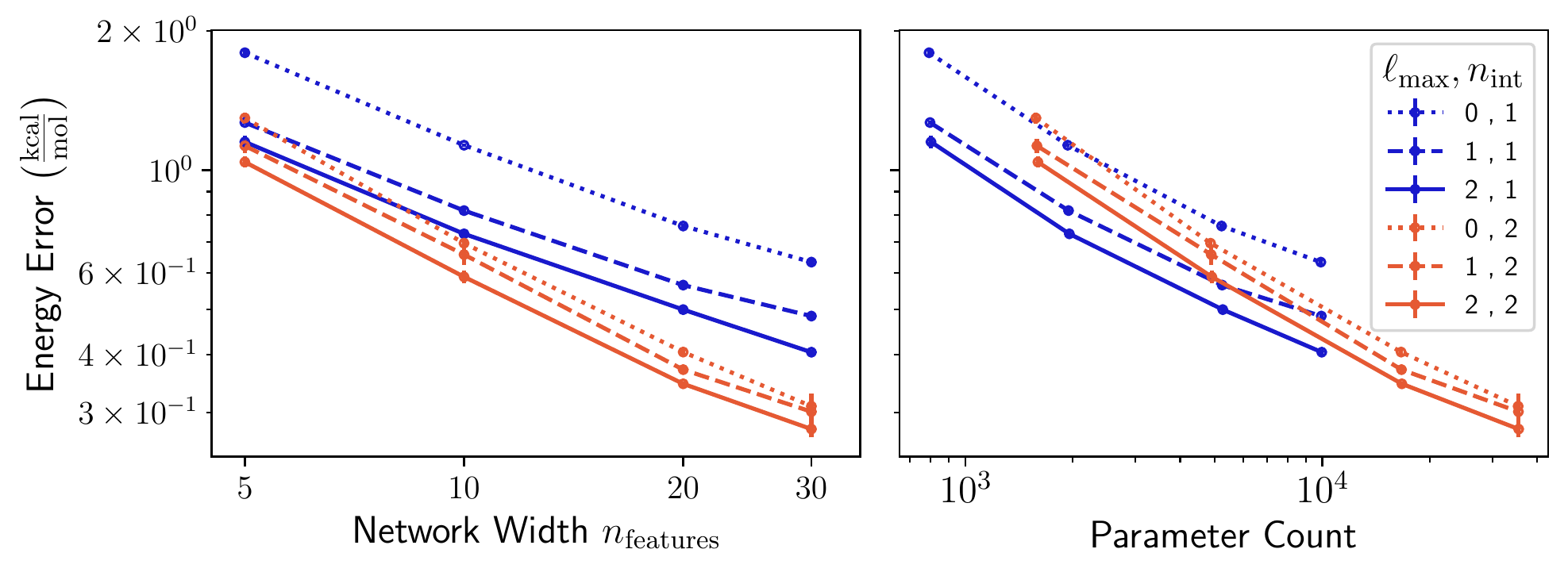}
    \caption{Mean absolute error of HIP-NN-TS models trained to QM9 for various network architectures. The original HIP-NN is a special case, with $\ell_\mathrm{max}=0$. Blue curves show models with one interaction block, red curves show models with two interaction blocks. Error bars (barely visible on some points) indicate the standard error of the mean over eight models. The left panel varies the number of artificial neurons per layer (network width), and the right panel shows the corresponding number of learnable parameters in the network. Increasing the tensor sensitivity order is helpful in all cases, but especially for shallower models. }
    \label{fig:qm9_small}
\end{figure*}

\textbf{Very small models of QM9}. The hyperparameter search of Table~\ref{tab:dataset_overview} leans towards very wide models (large $n_\mathrm{features}$), and it is reasonable to speculate that the improvements brought by $\ell_\mathrm{max}>0$ are a result of the large number of parameters in these models (> 100k); with so many parameters, perhaps any additional signal brought to the network allows for performance improvements. To investigate this, Fig.~\ref{fig:qm9_small} further illustrates HIP-NN-TS accuracies on the QM9 benchmark for models with a smaller number of parameters. We studied varying network widths ($n_\mathrm{features} \leq 30$), tensor sensitivity orders ($\ell_\mathrm{max}$) and number of interaction blocks ($n_\mathrm{int}$). For the cases tested, hyperparameters that lead to larger models yield more accurate models. A key feature of the HIP-NN-TS is that the number of model parameters is nearly independent of the choice of $\ell_\mathrm{max}$. If one seeks to maximize the model accuracy for a fixed parameter count, a larger tensor sensitivity order is preferable, with quadrupole order ($\ell_\mathrm{max} = 2$) being the largest we tested. There are diminishing returns, however, and asymptotic computational costs scale quadratically with $\ell_\mathrm{max}$.  Remarkably, the $n_\mathrm{int}=1$, $\ell_\mathrm{max}=2$, $n_\mathrm{features}=5$ model reaches an error of $1.19 \kcpm$ using just 801 parameters. All this demonstrates that even highly under-parameterized models (i.e. those with far fewer parameters than training points), whose parameter counts are far lower than conventional neural networks, still significantly benefit from the Tensor Sensitivity scheme with weight tying.

\subsection{QM7-X}

We also performed tests on QM7-X\cite{qm7x}, a systematically constructed dataset of 4.2M DFT calculations on 7K molecules. Because these conformations are non-equilibrium, we also include force in the loss function. The performance can be compared against Ref.~\onlinecite{spookynet}, which includes results for SchNet\cite{Schutt2017a}, a scalar architecture, PaiNN\cite{schutt2021equivariant}, a tensor architecture, and SpookyNet\cite{spookynet}, a tensor architecture that includes non-local contributions to its predictions. The results are shown in Table~\ref{tab:qm7x_performance}. While HIP-NN-TS does not set a new state-of-the-art, it performs comparably in energy prediction to PaiNN. This fact may be surprising, given that PaiNN strongly outperforms HIP-NN-TS on the QM9 dataset (c.f. Tab~\ref{tab:qm9_comparison}). This emphasizes that no single dataset can serve to completely characterize the performance of any given model architecture. 

\begin{table}
	\begin{tabular}{ l | l | l | l | l}
		\multicolumn{1}{c}{Model}    \\

		\toprule
		
		  \multicolumn{1}{r|}{Split}  & \multicolumn{2}{c|}{Test} & \multicolumn{2}{c}{Unk. Mols.}  \\
		  \multicolumn{1}{r|}{Quantity} & \multicolumn{1}{c|}{$E$} & \multicolumn{1}{c|}{$F$} & \multicolumn{1}{c|}{$E$} & \multicolumn{1}{c}{$F$} \\
		\midrule
		SchNet\cite{spookynet}  & 50.847 & 53.695 & 51.275 & 62.770  \\
		HIP-NN & 32.57(22) & 52.34(13) & 40.62(41) & 61.83(16)\\
		HIP-NN-TS, $\ell=1$ & 18.69(17) & 30.54(13) & 24.95(25) & 37.37(15) \\
		HIP-NN-TS, $\ell=2$ & 15.71(08) & 26.24(12) & 20.46(16) & 32.08(16) \\
		PaiNN\cite{spookynet}  & 15.691 & 20.301 &  17.594  & 24.161 \\
		SpookyNet\cite{spookynet} & 10.620 & 14.851 &  13.151 & 17.326 \\
		\bottomrule

 	\end{tabular}
	\caption{Mean absolute errors for energy $E$ (in meV) and forces $F$ (in $\frac{\mathrm{meV}}{\angstrom}$) for test splits (`Test') and unknown molecule splits (`Ukn. Mols.') from the  QM7-X dataset. Uncertainties in parentheses, $(\cdot)$, indicate the standard error of the mean for the last two digits of the error over a set of models trained with different train/test/validation splits and random initialization seeds.}
	\label{tab:qm7x_performance}
\end{table}

Research on equivariant networks has shown that the inclusion of tensor-valued activations can improve the slope of the learning curve for the model error as a function of the number of datapoints in the training set\cite{e3equivariant,Rackers_2023}. We studied this with QM7-X, varying the order of sensitivity and fraction of the dataset used for training, with a fixed-size validation set. The results, shown in Appendix Figure.~\ref{fig:qm7x_learningcurve}, indicate that indeed HIP-NN-TS models exhibit an improved learning curve compared to HIP-NN. Confirmation of this trend is interesting given that HIP-NN-TS is sensitive to tensors, but not equivariant, that is, does not incorporate tensor-valued activations into all layers of the network.

\subsection{Building a model for a general set of organic systems in non-equilibrium configurations}

To test the transferability and extensibility of tensor sensitivity in HIP-NN-TS, we trained several models to the ANI-1x dataset\cite{Smith2018LessLearning}. This training dataset includes about 4.9M conformations of organic molecules, which were sampled using an active learning procedure~\cite{Smith2018LessLearning}. Datasets collected using active learning are known to be effective for increasing the conformational and chemical diversity\cite{Zaverkin2022,Smith2018LessLearning}. We trained six HIP-NN-TS architectures to ANI-1x; the parameters of these span $\ell \in \{0,1,2\}$ and $n \in \{1,2\}$, using $n_\mathrm{features}=128$. Full training details are given in Appendix Section~\ref{sec:TrainingDetails}. For each architecture, we trained an ensemble of 8 models using independent, random splits of the data for train, validation, and test subsets.  We then test the models, both individually and as an ensemble average predictor, to examine performance on COMP6. This measures the model's ability to generalize to structures that are typically larger and more complex than those used in training, including probing intermolecular interactions, larger molecules such as tripeptides, and even a few small proteins. 

\begin{figure*}
    \centering
    \includegraphics[width=0.8\textwidth]{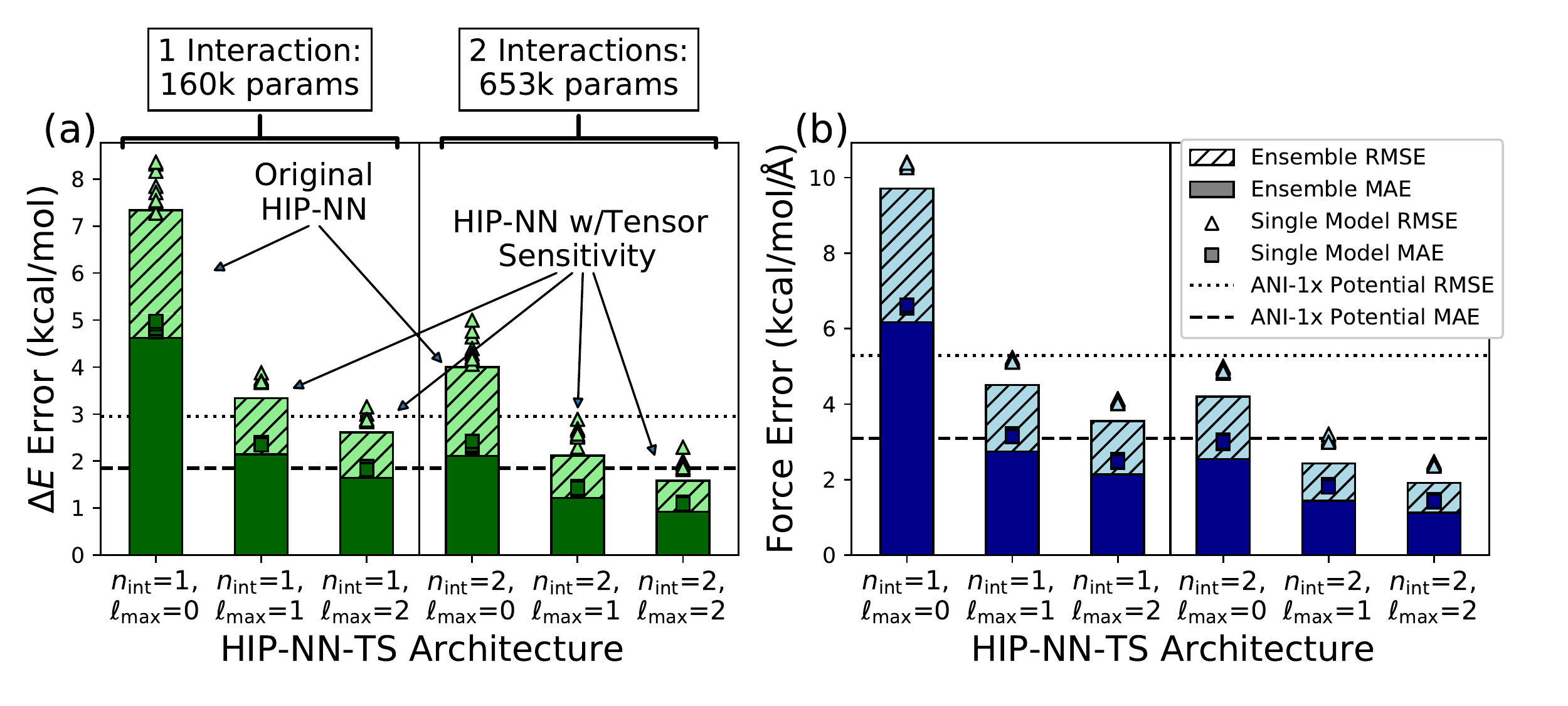}
    \caption{ML errors for ensemble models (bars) and single models (points) trained to ANI-1x and applied to COMP6. (a) Average ML errors for the predicted conformational energy, $\Delta E$. (b) Average errors of force components. More interaction layers $n_\mathrm{int}$ and higher orders in the tensor sensitivity ($\ell_\mathrm{max}=\{0,1,2\}$ for scalar, dipolar, and quadrupolar sensitivity) each yield improved performance.}
    \label{fig:comp6}
\end{figure*}

Figure~\ref{fig:comp6} provides errors for COMP6 tests in terms of both conformational energy $\Delta E$ and the force error $F$ in terms of both MAE and RMSE. Dashed lines show the performance of the ANI-1x potential~\cite{Smith2018LessLearning}, an ensemble of 5 ANI neural networks, which is a high-accuracy reference model well-known from the literature. From the figure, we observe that going beyond scalar sensitivities (i.e., setting $\ell_\mathrm{max} > 0$) consistently improves performance, as tested for models with $n_\mathrm{int} = 1$ and 2 interaction layers. As in the prior sections, diminishing returns are obtained in increasing from vector ($\ell_\mathrm{max}=1$) to quadrupole ($\ell_\mathrm{max}=2$) order. The benefit of setting $\ell_\mathrm{max} > 0$ is far more dramatic than previously observed in the QM7, QM9, and ANI1-ccx testing datasets; errors in both energy and force are reduced by a factor of 2 when moving from $\ell=0$ to $\ell=1$. This serves to demonstrate that model performance is strongly dataset dependent; the tensor sensitivity technique shows some improvement for QM9, but becomes far more important when applied to the challenging task of predicting the non-equilibrium conformations in COMP6. The improvements in the COMP6 performance are roughly similar to the improvements in the test set performance shown in Appendix Fig.~\ref{fig:extensibility_vs_test}. Ensemble models, which average over 8 HIP-NN-TS networks, consistently outperform single network predictions. Ensembling significantly reduces scatter in the energy predictions, yielding much lower energy RMSE, and a more marginal improvement on energy MAE and force errors. This suggests that individual networks are susceptible to occasional but significant misprediction of energies; these outlier predictions are mitigated through ensembling.

As Fig.~\ref{fig:comp6} shows, HIP-NN-TS compares favorably with previous ANI-1x benchmarks. Compared to the ensemble of 5 ANI-1x networks, a \textit{single} HIP-NN-TS network with $n_\mathrm{int}=1$, $\ell_\mathrm{max}=2$ performs similarly in energy accuracy, and significantly better in force accuracy. We also note that the HIP-NN-TS network architecture includes approximately $160\mathrm{k}$ parameters, whereas the full ANI-1x ensemble has $1.95\mathrm{M}$ total parameters for its 5 networks.

Even more striking is the performance of the two-interaction-block version of HIP-NN-TS. The ensemble of 8 HIP-NN-TS models ($n_\mathrm{int}=2$, $\ell_\mathrm{max}=2$) predicts conformation energies $\Delta E$ with an MAE of $0.927 \kcpm$, which is effectively half of the corresponding MAE for the ANI-1x ensemble. Even without ensembling, single HIP-NN-TS networks approach the $1 \kcpm$ chemical accuracy standard. For predictions of force components $F$, the HIP-NN-TS ensemble reaches $\mathrm{MAE}=1.24 \kcpmpa$, a factor of 2.75 lower than the ANI-1x ensemble.  It is worth noting that the ANI-1x ensemble did not train using force information; this may account for some improvement in force prediction. However, in preliminary tests, the effects of force training were not dominant; including forces in the loss for HIP-NN-TS improved the force prediction by roughly 35\%, and had mixed positive and negative effects on energy prediction within about 10\%.

\begin{table}
\centering
\begin{tabular}{l|c|c|c|c|c}
\toprule
 \multicolumn{1}{r|}{Split} & \multicolumn{1}{c|}{} & \multicolumn{2}{c|}{Full} & \multicolumn{2}{c}{<100 kcal}    \\

Model  & $f_\textrm{train}$ & $E$ & $F$ & $E$ & $F$  \\

\midrule

ANI-1x Ens.\cite{Smith2018LessLearning} & 1. & 1.85 & 3.09 & 1.61 & 2.70 \\

NewtonNet\cite{newtonnet} & 0.1 & - & - & 1.45 & 1.79 \\


HIP-NN ($n_\mathrm{int}=2$) Ens. & 1. & 2.57 & 2.55 & 2.24 & 2.14 \\

HIP-NN-TS ($n_\mathrm{int}=1$) Ens. & 1. & 2.02 & 2.14 & 1.74 & 1.86 \\

HIP-NN-TS ($n_\mathrm{int}=2$) & 0.9 & 1.26 & 1.44 & 1.06 & 1.23 \\

HIP-NN-TS  ($n_\mathrm{int}=2$)  Ens. & 1. & 1.06 & 1.12 & 0.88 & 0.95 \\

\bottomrule
\end{tabular}

    \caption{Performance of HIP-NN-TS and models from the literature on the COMP6 benchmark, using training to the ANI-1x dataset. Energy Errors $E$ are given in $\kcpm$, and Force Errors $F$ are given in $\kcpmpa$. The error metric is Mean Absolute Error (MAE). $f_\textrm{train}$ gives the fraction of the approximately 5M datapoints in the ANI-1x set that were used for training, including validation. `Full' indicates the entire COMP6 test set. The final set of columns indicates performance on conformations within 100 $\kcpm$ of the minimum energy conformer present in the database for each system. The HIP-NN-TS models in this table use $\ell_\mathrm{max}=2$.}
    \label{tab:ani1x_comparisons}
\end{table}

Performance details for each subset of COMP6 are shown in Appendix Table~\ref{tab:comp6_ensemble}, which demonstrates that the ensemble model performs well across all subsets. The single model performance exhibits similar trends with errors that are typically approximately 20\% higher, as recorded in Appendix Table~\ref{tab:comp6_singlemodel_errors}. A comparison with models from the literature is shown in Table~\ref{tab:ani1x_comparisons}, showing our results here compared with NetwonNet\cite{newtonnet}, who reported errors on a lower-energy subset of COMP6 and training to a fraction of the dataset, and the ANI-1x ensemble model\cite{Smith2018LessLearning}. HIP-NN-TS with one interaction layer provides comparable performance, and with two interaction layers, single models noticeably outperform NewtonNet, and the HIP-NN-TS ensemble even further improvements to accuracy. Another point of reference is the test of AIMNet~\cite{Zubatyuk2019AccurateNetwork}, which, being not precisely comparable, is not provided in the table. AIMNet was benchmarked to a variant of the Drugbank section of COMP6 containing more elements and computed at a different level of theory, and showed approximately a 25\% improvement over a corresponding ANI-architecture model. Based on the results in Appendix Table~\ref{tab:comp6_ensemble}, which show that HIP-NN-TS out-performs ANI by more than a factor of 2 (Drugbank ANI-1x: 2.09 $\kcpm$ MAE vs. HIP-NN-TS: 1.002 $\kcpm$ MAE), it is reasonable to conjecture that HIP-NN-TS would, in a head-to-head comparison, outperform AIMNet. 

We also performed analysis on the hierarchical nature of the predictions of HIP-NN-TS on COMP6 configurations, shown in Appendix Figure~\ref{fig:hierarchicality}, comparing the hierarchical parameter $R$ and the energy error, both directly and using a per-atom normalization. The results show that for higher order $\ell$, HIP-NN-TS is able to form more hierarchical predictions (lower $R$). There is a correlation between $R$ and error, as observed previously in Ref.~\onlinecite{Lubbers2018}. However, on a per-atom basis, this relationship is diminished; much of the correlation can be explained by the system size.

\subsection{Computational Throughput of HIP-NN-TS}

Finally, we remark on and perform studies of the trade-off between improved accuracy and increased computational complexity of inference between HIP-NN, HIP-NN-TS, and other models in the literature. While the flops required for tensor sensitivity scale quadratically with asympotically large $\ell_\mathrm{max}$, in practice, the values $\ell_\mathrm{max} = 0,1,2$ are low enough that other computational costs remain significant. Our most accurate $\ell_\mathrm{max}=2$ models of ANI-1x exhibited training times of approximately 830 s/epoch on an NVIDIA A100 GPU, which is approximately 2 times slower than corresponding $\ell_\mathrm{max}=1$ models, and 4 times slower than the counterpart without tensor sensitivity, $\ell_\mathrm{max}=0$. This translates to a training throughput on the order of 20 $\upmu$s per atom evaluation in training. The expected performance in deployment is somewhat context-dependent. To reach maximum GPU efficiency, large batches of approximately 8k atoms per network evaluation are desirable. To achieve this GPU saturation for simulations of small systems, one could simulate multiple copies of the same system in tandem~\cite{Zhou2020}. Typical simulations will not need to compute gradients of the loss function, which saves an approximate factor of two in cost relative to training.
Training of all models, including our largest networks that address ANI-1x and QM7-X, was accomplished within a highly practical two day wall-clock time on a single A100 GPU, and the per-atom evaluation estimate for the largest model appears competitive with other recent techniques~\cite{Zuo2020, Lysogorskiy2021}. The one-layer $n_\mathrm{int}=1$, $\ell_\mathrm{max}=2$ models trained approximately four times faster than those with $n_\mathrm{int}=2$. There are 32 times more features entering the second interaction layer than the first interaction layer, so the theoretical flop count to apply the weights of the second layer is 32 times greater. The observed factor of four slowdown going from $n_\mathrm{int}=1$ to 2 is therefore relatively small. This smallness illustrates that additional portions of the model play a role (e.g., the costs to calculate the pairwise sensitivities).

It is also important to consider the throughput of HIP-NN in comparison to other models from the literature, for which some results have been reported. For this test, performing inference on QM9 at batch size 50, we used a network architecture identical for the training to the ANI-1x dataset. The results are shown in Table~\ref{tab:qm9_speed}. However, the time for a batch size of 50 is dominated by overheads, and so we also probe larger batch sizes which are more able to effectively saturate the GPU. While only 40\% faster than the Equivariant Transformers (ET) model\cite{tholke2022torchmd} at batch size 50, HIP-NN-TS with $\ell=2$ is approximately 5 times faster than the reported ET results when scaling to large batch sizes. Spending 29 $\upmu$s per molecule prediction, HIP-NN-TS with $\ell=2$ can perform inference on the entire QM9 dataset in less than five seconds, and HIP-NN-TS with $\ell=1$ is twice as fast. While fast, this is still almost 3 times slower than scalar HIP-NN, as shown in the table. We emphasize again that this a relatively large model; smaller models (such as those presented in Sec.~\ref{sec:qm7qm9ani1ccx}) may provide adequate performance with higher throughput.

\begin{table}
\centering
\begin{tabular}{l|c|c|c}
\toprule
 & Batch & Parameter & Time per  \\
 Model  & Size & Count & Molecule ($\upmu$s) \\

\midrule
DimeNet++\cite{schutt2021equivariant} & 50 & 1.8M & 900 \\
PaiNN\cite{schutt2021equivariant} & 50 & 600K & 250 \\
ET-large\cite{tholke2022torchmd} & 50 & 6.87M & 234 \\
ET-small\cite{tholke2022torchmd} & 50 & 1.34M & 188 \\
HIP-NN-TS,\,$\ell=2$ & 50 & 656K & 116 \\
HIP-NN-TS,\,$\ell=1$ & 50 & 656K & 90 \\
HIP-NN & 50 & 656K & 76 \\
HIP-NN-TS,\,$\ell=2$ & 512 &656K  & 29 \\
HIP-NN-TS,\,$\ell=1$ & 1024 &656K  & 15 \\
HIP-NN & 2048 & 656K & 6.7 \\

\bottomrule
\end{tabular}

    \caption{Inference throughput of models applied to QM9. Timings given in terms of $\upmu$s per system energy inference; lower is better.}
    \label{tab:qm9_speed}
\end{table}

We also trained HIP-NN and HIP-NN-TS to the Allegro-Ag\cite{allegro} dataset and performed large-scale molecular dynamics.  Allegro is a unique and interesting architecture focused on the processing of pair-valued neural network variables within a fixed cutoff radius, that is, without using a message-passing. The Allegro-Ag dataset was constructed from an \textit{ab initio} MD run of an FCC structure with a vacancy, and, due to its comparatively simple nature, allows probing the limits of accuracy and throughput for a single-phase, single-element dataset. Training of all three models ($\ell \in \{0,1,2\}$) was performed concurrently on a laptop CPU, and completed on this relatively small dataset in under two hours. The HIP-NN-TS models were able to learn to comparable accuracy to the corresponding Allegro model, as shown in Appendix Table~\ref{tab:allegro_ag_performance}, using less than one tenth of the parameter counts, and the accuracy of unmodified HIP-NN is only about 20\% worse than this. These models were set to reflect the interaction ranges used by the Allegro model, with a hard cutoff at $4 \angstrom$. The Allegro model used the analog of one interaction layer, and $\ell_\mathrm{max}=1$.

Molecular Dynamics (MD) was performed using the LAMMPS\cite{lammps_2022} MLIAP package for Machine Learning Interatomic Potentials, in particular, using the recently available ``MLIAP Unified pair style'' feature of LAMMPS, which provides for end-to-end Deep Neural networks implemented in Python to be used in LAMMPS, including all-GPU MD interfacing with LAMMPS Kokkos\cite{kokkos} arrays; this implementation is generally comparable with the custom Kokkos implementation for Allegro.

\begin{figure*}
    \centering
    \includegraphics[width=0.7\textwidth]{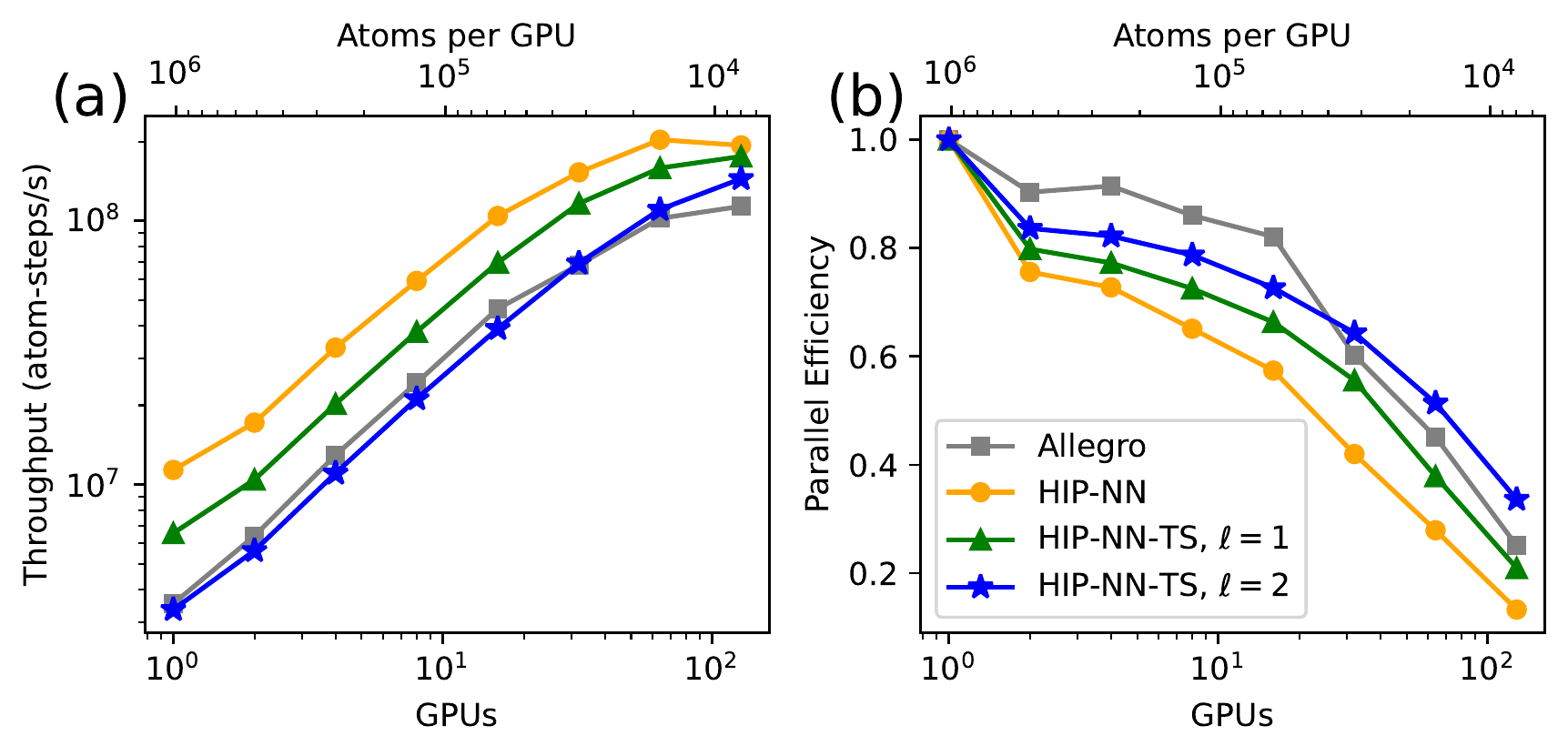}
    \caption{Molecular Dynamics (MD) performance of HIP-NN-TS compared with the results of the Allegro model experiments\cite{allegro}. Models were trained on the Allegro-Ag dataset, and MD performed on a 1,022,400 atom system. (a) Throughput of the models in strong-scaling, that is, apply the model to a fixed system size across a variable number A100 GPUs. (b) Parallel efficiency, that is, the ratio of speedup to resources compared to a single-GPU run. For both metrics, larger values are preferable.} 
    \label{fig:allegro_md}
\end{figure*}

 We then performed a strong-scaling study of a 1M atom FCC system as described in Ref.~\onlinecite{allegro}, varying the number of GPUs from 1 to 128. The results of this study are shown in Figure~\ref{fig:allegro_md}. The HIP-NN-TS ($\ell=2$) model exhibits a very similar computational throughput and parallel efficiency to Allegro, and strong-scales slightly better to a large number of GPUs. The HIP-NN-TS ($\ell=1$) model exhibits better throughput by almost a factor of two, with slightly worse strong-scaling efficiency. HIP-NN (without tensor sensitivity) has higher throughput but does not strong-scale effectively to systems with less than $10^4$ atoms; the original HIP-NN is in some sense too fast to achieve maximum GPU efficiency for small systems. Taken in sum, this study shows that HIP-NN-TS performs somewhat better than Allegro on this test, pushing forward the frontier for lightweight, speedy, and accurate models for metallic systems in a known single phase.

\section{Discussion}

In all cases tested, using higher order tensor sensitivity, controlled by $\ell_\mathrm{max}$, improves HIP-NN-TS accuracy. The magnitude of improvement, however, depends on the benchmark. We observe the biggest performance improvements on the challenging datasets that have a large conformational diversity, such as ANI-1x and COMP6. For datasets without conformational diversity, such as QM7 or QM9, the benefits of increasing $\ell_\mathrm{max}$ were more marginal. This result is in agreement with the ablation study of the authors of Equivariant Transformers, which compared $\ell=0$ and $\ell=1$ models, and found a 6\% performance improvement on QM9, and a 146\% performance improvement on nonequilbrium molecular trajectories~\cite{tholke2022torchmd}, although the effect was less pronounced in our tests.  Furthermore, our survey of the ML-potential literature for QM9 did not indicate any clear pattern between geometric depth and tensor order hyperparameters, and the resulting model accuracy. A useful feature of HIP-NN-TS is that the count of model parameters is nearly independent of the choice of $\ell_\mathrm{max}$, which makes it possible to study the effects of tensor order without introducing confounding factors; one-to-one comparison is straightforward.

Two aspects of the tensor message mixing formula in Eq.~(\ref{eq:hip-vec-1}) merit further discussion:
First, the monopole term $\mathcal{I}_{i,a}^{\textrm{HIP-NN}}=\mathcal{E}_{i,a}^{(0)}$
retains its sign, whereas no sign is present for the tensor norms
that appear at orders $\ell\geq1$. The choice to retain the sign at order $\ell = 0$ ensures that HIP-NN-TS is a strict generalization of HIP-NN.
A second aspect of Eq.~(\ref{eq:hip-vec-1}) is that the Frobenius norm $|\cdot|$
introduces a cusp (conical) singularity at zero.
We made this choice so that the model parameters $v_{ab}(r_{ij})$ appearing in Eq.~\eqref{eq:m_def}, would contribute linearly to $A^\textrm{HIP-NN-TS}_{i,a}$ at each tensor order $\ell$.  We provide a discussion of alternatives that avoid this cusp in Appendix Section~\ref{sec:alternative_ts}.

We also remark on a fundamental limit to the information capacity of tensor sensitivity. The tensor norms appearing in the HIP-NN-TS interaction rule, Eq.~(\ref{eq:hip-vec-1}), can each be expressed as
\begin{equation}
|\boldsymbol{\mathcal{E}}_{i,a}^{(\ell)}|  =\Big|\sum_{j}\mathbf{M}_{i,j,a}^{(\ell)}\Big|
 =\sqrt{\sum_{j,j'}\mathbf{M}_{i,j,a}^{(\ell)}\cdot\mathbf{M}_{i,j',a}^{(\ell)}},
\end{equation}
where the dot denotes contraction on all tensor indices. The crucial
thing to observe is that only \emph{pairwise} coupling between messages
from $(j,j')$ to $i$ is possible. In particular, the descriptors
constructed in a single HIP-NN-TS interaction layer will miss information
from the bi-spectrum~\cite{GAPoriginal,AtomicClusterExpansion} and are subject to all the limitations discussed in Ref.~\onlinecite{Pozdnyakov2020}. In particular, it remains possible to construct geometric configurations that cannot be distinguished from each other within a single interaction layer, regardless of $\ell_\mathrm{max}$. In practice,
HIP-NN and related message passing models overcome such limitations
by using multiple interaction layers, allowing reconstruction of arbitrary higher order $N$-body information~\cite{Tasissa2018, Glunt1993}. 

Work along this line has been examined using BOTNet\cite{batatia2022design} and MACE\cite{batatia2022mace} models, which have explored generalized versions of the NEquip\cite{e3equivariant} tensor architecture by incorporating elements of ACE\cite{AtomicClusterExpansion}. We now remark on the three main differences between HIP-NN-TS and these and other approaches.

First, the form of HIP-NN is different from SchNet-like models, including NEquip. Where HIP-NN and HIP-NN-TS use a matrix-valued convolution as given in Eq.~(\ref{eq:a_hipnn_def}), which closely mimics the form of Convolutional Neural Networks used in image processing\cite{lecun2015deep}, these other models use a factorized form based on a Hadamard product of a feature transformation with a radial envelope function. In a very heuristic phrasing, HIP-NN messages are allowed to mix across the feature dimension during propagation over distance, whereas Schnet-like messages can have modulated amplitude, but cannot mix while propagating. The form of HIP-NN is more similar to the feature-and-distance outer product used in Behler-Parinello\cite{Behler2015} networks, ANI\cite{Smith2017b}, and AIMNet\cite{Zubatyuk2019AccurateNetwork}. One can conjecture that this increased flexibility explains how HIP-NN and HIP-NN-TS are able to address complex datasets with only one to two interaction layers. More remarks on this can be found in Ref.~\onlinecite{extendingbeyondproperties}. Both forms of scalar interaction fit into the general framework of a message passing network as defined in Ref.~\onlinecite{Gilmer2017} and the Multi-ACE framework\cite{batatia2022design}. In the Multi-ACE terminology, SchNet-like messages do not include pooling weights. The HIP-NN form corresponds to a factorized choice of the pooling weights $w$ identified by of Multi-ACE. The full structure of Multi-ACE also incorporates the possibility received messages on an atom $i$ depend on the features already present on atom $i$, where in HIP-NN, these features are incorporated separately from the message-passing procedure. 

Secondly, HIP-NN-TS uses a simpler form of tensor mathematics than works deriving from Tensor Field Networks\cite{Thomas2018}, including NEquip, BOTNet, and MACE, in two ways. HIP-NN-TS does not mix tensor channels using tensor product reduction but rather extracts scalar values from each tensor order and mixes these scalars together. Several equivariant models, such as Cormorant and Tensor Field networks, NEquip, etc, use the spherical representation for geometric tensors. In these models, it is necessary to use Clebsch-Gordon coefficients to compute the tensor product decomposition into a sum of invariant representations. The Cartesian representation employed here involves only elementary operations on real numbers, and we have found it  easy to implement in PyTorch. Cartesian tensors also facilitate a natural interpretation in terms of the multipole expansion: $\ell = 1$ describes the dipolar character of the messages from the local atomic environment, $\ell = 2$ the quadrupolar character, and so on. The choice to use Cartesian tensors rather than spherical harmonics is, however, an implementation detail; a given model could be described in either language. HIP-NN-TS also uses weight tying---the same parameters $v_{ab}(r_{ij})$ are reused for every $\ell$, which represents an inductive bias towards a message passing form which, on the feature dimension, does not vary between tensor orders. The total environmental interaction for a given atom, Eq.~(\ref{eq:hip-vec-1}), can be interpreted a weighted combination of these multipolar parts. This is why HIP-NN-TS contains very few extra parameters in comparison to HIP-NN. At the same time, because of the simplified tensor mathematics, HIP-NN-TS as presented here does not produce equivariant predictions (only invariant predictions) at arbitrary tensor order. Equivariant predictions could, in future work, be constructed analogously to HIP-NN-TS by generalizing the quadratic form of bond order predictions with HIP-NN\cite{magedovbond} to include tensor information.

Thirdly, in this work we have used low-order tensor invariants because they capture three-body terms relatively efficiently in computational cost, and extremely efficiently in parameters added to the model. Higher order tensor sensitivities highlighted in Multi-ACE\cite{batatia2022design} are possible, which could take into account more complex tensor invariants, along the lines of the ACE~\cite{AtomicClusterExpansion} or Moment Tensor Potentials~\cite{Shapeev2015MomentPotentials},
and message passing neural networks using generalized ACE descriptors~\cite{batatia2022design}. Such schemes allow the incorporation descriptors at the $n$-body level, but the count of descriptors grows rapidly with $n$. Cost is a concern, as the computational cost of MACE was characterized with a latency of 10ms per prediction on a single molecule using only scalar message passing\cite{batatia2022mace}, where HIP-NN-TS with $\ell=2$ has a prediction time of less than 6ms even on batches 50 molecules (c.f. Tab.~\ref{tab:qm9_speed}). An intriguing possibility is to explore construction of {\it independent} invariants, which can be fewer in number~\cite{bujack2017moment}; this would allow a network to profit from the many-body information in higher order tensor invariants at significantly reduced computational cost. 

At the same time, one must consider the role of inductive bias\cite{baxter2000model} in data-driven modeling, which is the primary route for understanding how data-driven models can be effective in light of the No Free Lunch theorems\cite{wolpert1996lack}. The atomic cluster expansion formalism\cite{AtomicClusterExpansion}, and models invoking it, use a formally complete expansion within a given local basis set; the only inductive bias is locality and choice of local basis set. Thus, all other architectures correspond to choices of some ansatz about the coefficients of the infinite series. The situation can be analogized to wavefunction-based approaches in \textit{ab initio} quantum mechanics, where the full wavefunction may include arbitrary $n$-body electron correlations, leading to the exponential costs of Full Configuration Interaction. Approximations to the wavefunction such as Coupled-Cluster theory explore a reduced set of possible correlations that nonetheless capture many phenomena of interest. However, data-driven modeling contains an important additional twist, because the determination of the infinite series of coefficients in the atomic cluster expansion requires an infinite amount of data to constrain the model. The inductive bias incurred by any particular ansatz can be useful to locate better-generalizing approximations with a finite amount of data that lies on a manifold. As a concrete example of this phenomenon, consider that HIP-NN and HIP-NN-TS both out-perform the ANI network architecture on QM9, a dataset of equilibrium conformations. However, in the transferability tests to COMP6, HIP-NN performs worse than ANI, where HIP-NN-TS performs far better. 

\section{Conclusions and Future Work}

We introduced HIP-NN-TS, which extends HIP-NN with Tensor Sensitivities. This generalization provides direct and clear benefits in terms of accuracy as shown on the QM7, QM9, QM7-X ANI-1ccx, COMP6, and Allegro-Ag datasets. Through the use of weight tying between the sensitivities at all tensor orders $\ell = 0, \dots \ell_\mathrm{max}$, this scheme requires only one additional hyperparameter, the maximum order $\ell_\mathrm{max}$. The only new model parameters are scalars $t_a^{(\ell)}$ that describe the strength of mixing between the different tensor orders; typically, these will be fewer than 1\% of the total number of model parameters. Tensor sensitivity is implemented as a type of spectral operator, rather than an explicit $N$-body calculation, and as such does not imply heavy scaling in dense environments, as is found in explicit angular models.

We also studied the computational throughput for HIP-NN-TS, both for inference of small molecules in QM9, and in strong-scaling parallel molecular dynamics of 1M silver atoms. The performance of HIP-NN-TS is somewhat slower than scalar HIP-NN (a factor of 2 to 5, depending on the setting), but is still fast in comparison to several models in the literature that also provide high accuracy. Overall, these studies highlight the need to explore the trade-off between error and computational efficiency, both in choosing the network architecture, and in choosing the architectural hyperparameters, in any given application.

Although model parameter count remains approximately constant,  the number of activations scales approximately quadratically with tensor order $\ell_\mathrm{max}$. This hyperparameter should therefore be selected with memory and computational performance in mind. Despite this additional cost, it is still feasible to train to very large datasets, such as ANI-1x (5 million small molecules) and QM7-X (4.2 million small molecules), within 48 hours on a single A100 GPU. The results given here might be improved upon by using longer training runs. At the same time, for some modeling tasks, the improved performance of tensor sensitivity might not be worth the additional computational expense. For example, in active learning workflows~\cite{Smith2018LessLearning}, the training time provides constraints on the time between generations of data collection. 

Future work could take several directions. There are innumerable applications of ML-based potentials and models to explore. One could also incorporate more physics into the model architecture, such as estimation of charge and Coulombic interactions~\cite{Unke2019PhysNet:Charges,Ko2021}, that may lead to improved transferability and performance.
Non-local machine learning terms, such as were investigated by SpookyNet\cite{spookynet}, are known to be needed in order to capture certain quantum-mechanical phenomena (such as conjugation in organic systems) that are themselves not local. On the other hand, non-local terms may present difficulties for domain decomposition techniques used in parallel simulation. These techniques might extend the capabilities of HIP-NN-TS. Another direction for improvement of the model is to consider using higher-body-terms for the messages used in each interaction layer; this would address known limitations of information capacity for single neurons that capture only 3-body information~\cite{Pozdnyakov2020}. Hierarchical analysis with respect to the variable $\ell$ has recently been examined\cite{rackers2022hierarchical} for other models, and it would be interesting to include both depth and tensor order in hierarchical regularization.

Finally, we remark on some intriguing results that appeared in our hyperparameter searches. For the very small QM7 bechmark dataset, one interaction block performed better than two. But in all benchmarks that we explored, the widest models tested were the best performing. This implies that overfitting is more likely to occur as a result of increasing depth rather than increased width, which could be interesting to study in future work, in which one could examine the trade-offs involved in models with large $n_\mathrm{feature}$, asking at at what point are the diminishing returns of increased accuracy worth the computational cost of increasing the width of the network. One may also investigate whether the difficulty with finding models that perform better with more than two interaction blocks is fundamental, or a result of other factors, such as hierarchical regularization or the optimization algorithm.  

Code that implements HIP-NN-TS using PyTorch~\cite{NEURIPS2019_9015} is available within the \texttt{hippynn} Python package, available at \url{github.com/lanl/hippynn}. An included example file demonstrates training to ANI-1x. All training datasets used in this study are available publicly~\cite{Rupp2012, Ramakrishnan2014QuantumMolecules, Smith2020TheMolecules,qm7x,allegro}.

\acknowledgments

The authors would like to acknowledge Alice E. A. Allen and Sakib Matin for valuable discussions and feedback.

This project benefited from computing resources provided by the CCS-7 Darwin cluster, funded by the Computational Systems and Software Environments (CSSE) subprogram of ASC at LANL. Additional computing resources were provided by the Center for Integrated Nano Technologies (CINT) at LANL. This research used resources provided by the Los Alamos National Laboratory Institutional Computing Program, which is supported by the U.S. Department of Energy National Nuclear Security Administration under Contract No. 89233218CNA000001. M. Chigaev and J. S. Smith acknowledge support from the Center for Nonlinear Studies (CNLS) and the Laboratory Directed Research and Development Program at LANL. S. A. was supported by the U.S. Department of Energy, Office of Fusion Energy Sciences (OFES) under Field Work Proposal Number 20-023149.
K. Barros, B. Nebgen, and N. Lubbers acknowledge support from the US DOE, Office of Science, Basic Energy Sciences, Chemical Sciences, Geosciences, and Biosciences Division under Triad National Security, LLC (“Triad”) contract Grant 89233218CNA000001 (FWP: LANLE3F2).

\appendix

\section{Dataset Details}
\label{sec:DatasetDetails}

\subsection{QM7}
The QM7 dataset~\cite{Rupp2012} includes approximately 7k organic molecules (a subset of nearly 1 billion molecules within GDB-13~\cite{BlumGDB13}) containing the elements \{H, C, N, O, S\} with up to 7 heavy (non-H) atoms. The molecular geometries are relaxed to minimize the DFT conformational energy~\cite{Rupp2012}. The dataset contains many structures such as double and triple bonds, amides, alcohols and others~\cite{Rupp2012}. The Cartesian coordinates were generated with OpenBabel using strings from GDB-13. The training target is atomization energy, which was generated using the Perdew-Burke-Ernzerhof hybrid functional (PBE0) approximation~\cite{Rupp2012}.

\subsection{QM9}
The QM9 dataset~\cite{Ramakrishnan2014QuantumMolecules} describes approximately 130k organic molecules of elements \{H,C,N,O,F\} using up to 9 heavy atoms with relaxed geometries. The dataset supplies many molecular properties, but our focus is atomization energy, calculated at the B3LYP/6-31G(2df,p) level of quantum theory. About 3k molecules were flagged due to failing a geometric consistency check, and we removed them from our training dataset.

\subsection{ANI datasets: ANI-1x, ANI-1ccx, and COMP6}

The ANI-1x dataset~\cite{Smith2020TheMolecules} consists of approximately 5M DFT ($\omega$b97x, 6-31G(d) method) calculations from approximately 64k distinct organic systems in various non-equilibrium conformations. Molecules in this set contain only the elements \{H,C,N,O\}.
The geometries were produced using an iterative active learning strategy, which seeks molecular conformations for which there is a large model uncertainty~\cite{Smith2018LessLearning}. This uncertainty was estimated using the disagreement between five neural network models constituting an ensemble.
Candidate conformations were sampled using a variety of techniques, including distillation of the prior ANI-1 dataset, normal mode sampling, and molecular dynamics on both single molecules and limited number of dimer systems.

A related dataset is ANI-1ccx\cite{Smith2019ApproachingLearning,Smith2020TheMolecules}, which includes about 490k geometries (about 10x fewer than ANI-1x), calculated using coupled cluster theory, at the approximate level of CCSD(T)/CBS. The molecules contain only \{H, C, N, and O\}, with conformations primarily downselected from ANI-1x using an active learning procedure. ANI-1ccx additionally includes dihedral sampling of small molecules not present in ANI-1x.  The number of systems in ANI-1ccx was not reported. If 10\% of the data from ANI-1x had been randomly selected, virtually all of the systems should be present in ANI-1ccx, because ANI-1x contains on average 76 configurations per system. Since active learning encourages diversity in selection, it seems even more likely that nearly all systems in ANI-1x would be found in ANI-1ccx. As such we label it as having 64K chemical systems in table~\ref{tab:dataset_overview}. Still, this is an estimate.

An important test of the models trained to ANI-1x is the COMP6 benchmark, which is computed at the same level of theory as ANI-1x~\cite{Smith2018LessLearning}. The COMP6 benchmark consists of 6 subset molecular groups: drugbank molecules, tripeptides, s66x8 (an existing benchmark for noncovalent interactions), GDB7-9 (containing molecular conformations from molecules in the GDB13 set which have between 7 and 9 heavy atoms), GDB10-13, and ani\_md (containing well-known drug molecules and two proteins with conformations obtained by molecular dynamics using ANI-1x).  These groups represent a diverse set of tests, designed to comprehensively assess model error on benchmarks that are more challenging than the ANI-1x training set itself; the test molecules are out of sample and far larger than those present in the training data.

\subsection{QM7-X}

QM7-X\cite{qm7x} is a systematically constructed dataset of 4.2M DFT calculations on 7K molecules including \{H, C, N, O, S, Cl\}. For each molecule of interest, stationary points in the potential energy surface were located using Density Functional Tight Binding. Around each of these minima, 100 non-equilibrium conformations were identified using a diverse normal-mode-sampling algorithm. The resulting conformations were evaluated using Density Functional Theory including a many-body dispersive correction scheme.

\section{Model and training details}
\label{sec:TrainingDetails}

The training procedure in this work follows closely with the methods in Ref.~\onlinecite{Lubbers2018}. For both HIP-NN and HIP-NN-TS, we employed the loss function,
\begin{equation}
\mathcal{L} = \lambda_\mathrm{error} (\mathrm{RMSE} + \mathrm{MAE}) + \mathcal{L}_{\mathrm{L2}} + \mathcal{L}_{R}. \nonumber
\end{equation}
The loss function includes two types of error (MAE and RMSE), as well as regularization terms. The first regularization term $\mathcal L_{\mathrm{L2}} = \lambda_{\mathrm{L2}} \cdot (\dots)$ applies an L2 (squared magnitude) penalty on all model parameters (weights and biases). The second regularization $\mathcal L_{R} = \lambda_R \langle R \rangle$ penalizes the non-hierarchicality $R$ as defined in the original HIP-NN work, and is averaged over all points in the dataset. For the QM7, QM9, and ANI-1ccx datasets, we weight the loss function terms according to the hyperparameters
\begin{equation*}
\lambda_{\mathrm{L2}} = 10^{-6},\quad\lambda_{\mathrm{error}} = \frac{1}{(230 \mathrm{kcal}/\mathrm{mol})},\quad \lambda_R = 10^{-2}.
\end{equation*}
as in prior work~\cite{Lubbers2018}. For ANI-1x, we use
\begin{equation*}
\lambda_{\mathrm{L2}} = 10^{-6},\quad\lambda_{\mathrm{error}} = 1 ,\quad \lambda_R = 1,
\end{equation*}
and an additional prediction error term is added to the loss to penalize MAE and RMSE for the force components.
For QM7-X, we use
\begin{equation*}
\lambda_{\mathrm{L2}} = 10^{-6},\quad\lambda_{\mathrm{error}} = 1 ,\quad \lambda_R = 2\times 10^{-3},
\end{equation*}
with a force weighting of 10; we observed a similar but weaker effect than SpookyNet\cite{spookynet}, who observed strong benefits at higher force weighting of 100. In the case of HIP-NN-TS, we found that a force weight of 10 provided approximately a 10\% benefit to energy errors while still improving on force errors. For Allegro-Ag, we use
\begin{equation*}
\lambda_{\mathrm{L2}} = 10^{-6},\quad\lambda_{\mathrm{error}} = 1 ,\quad \lambda_R = 1,
\end{equation*}
and an energy weighting of 10 compared to forces.

HIP-NN is characterized by three main hyperparameters defining the model depth and width. There are two types of layers: (1) \emph{on-site} (\emph{atom}) layers, which process information only \emph{within} a single site, and (2) \emph{interaction} layers,
which pass information between neighboring sites. Both are described by Eq.~(\ref{eq:hipnn-layer}), but for on-site layers, we set $v_{a,b}^{\ell}(r_{ij})=0$. 
We perform several experiments with different model sizes, characterized by the number of interaction layers $n_\mathrm{interactions}$, the number of on-site layers $n_\mathrm{layers}$ following each interaction layer, and the layer width, $n_\mathrm{features}$. As expected, the number of model parameters grows linearly in $n_\mathrm{interactions}$ and $n_\mathrm{layers}$, but quadratically in $n_\mathrm{features}$. HIP-NN-TS introduces an additional hyperparameter $\ell_\mathrm{max}$, which controls the tensor order of sensitivities ($\ell=\{0,1,2\}$ describe scalar, vector, and quadrupole sensitivities, respectively, and in general contain $2\ell+1$ components). Messages passed between atoms include a total of $\sum_{\ell=0}^{\ell_\mathrm{max}} (2\ell + 1) = (\ell_\mathrm{max}+1)^2$ components. We performed an exhaustive hyperparameter search to find the optimal $n_\mathrm{int}$ and $n_\mathrm{layers}$, whereas the choice of $n_\mathrm{features}$ and $\ell$ is typically made as a tradeoff between model accuracy and efficiency.

Equation~(\ref{eq:hipnn-layer}) takes as inputs the feature vectors $z_{j}$ (for all sites $j$, still fixing layer $\ell$) and yields as outputs $\tilde{z}_{i}$. A subsequent ResNet transformation
on $z_{i}$ and $\tilde{z}_{i}$ then generates the feature vectors
of the next layer. We employ
the softplus activation function $f(x) =\log(1+e^{x})$, where $x$
is a scalar.

In training, we use the Adaptive Moment Estimation algorithm (Adam)~\cite{Kingma2014Adam:Optimization}, which is a variant of the Stochastic Gradient Descent Algorithm (SGD). The goal is to find the set of model parameters $\Theta$ that minimize the total loss $\mathcal{L}$ on a validation dataset. SGD splits the training data into batches, and then changes $\Theta$ by evolving it in the direction of the negative gradient of the stochastic approximation of the loss. The algorithm does this in epochs, which is when the NN does a full pass through the dataset. Adam helps speed convergence by evolving $\Theta$ through a decaying average of previous gradients. There are three parameters in the Adam algorithm, the learning rate $\eta$, and exponential decay factors $\beta_1 = 0.9$ and $\beta_2 = 0.999$. In order to reduce overfitting, a patience/early stopping system is used. When training, the learning rate is initialized to $\eta_\mathrm{init} = 10^{-3}$ for QM7, QM9, ANI1-ccx, and Allegro-Ag, and $\eta_\mathrm{init} = 5\times10^{-4}$ for ANI-1x and QM7-X, which improves model convergence when using a smaller patience parameter.  The best score, which is the lowest MAE validation value observed, is tracked. The learning rate is fixed for at least $n_{\textrm{patience}}$ epochs, to be defined below. If the best score does not improve during this time, the learning rate is multiplied by a decay factor $\alpha = 0.5$, which decreases the learning rate and causes the training procedure to perform finer adjustments to the weights. If the model improves, the patience counter is reset, running again for at least $n_{\textrm{patience}}$ epochs.  Training to QM7, QM9, and ANI-1ccx datasets used $n_{\textrm{patience}}=50$ epochs, while training to the much larger ANI1x and QM7-X used $n_{\textrm{patience}}=5$ epochs. Training to Allegro-AG used $n_{\textrm{patience}}=25$ The annealing schedule follows the procedure of Ref.~\onlinecite{SamuelSmith2018}, which raises the batch-size from a beginning value to a final value before lowering the learning rate. For QM7 and QM9 the batch size was varied between 256 and 1024. For ANI-1ccx the batch size was varied between 512 and 2048. For ANI-1x, a fixed batch sized of 512 was used. Allegro Ag used a batch size of 4. Training is stopped if decreasing the learning rate twice yields no improvements, or if a maximum runtime is reached: QM7, QM9, and ANI-1ccx used a maximum of 2000 epochs, which in practice was not reached. Training to Allegro used a maximum of 400 epochs, ANI-1x used a maximum of 150 epochs, and QM7-X used a maximum of 130 epochs. Models of QM7-X sometimes stopped early as a result of the low patience, affecting train, validation and test errors alike, and so only models which reached at least 100 epochs were used in the final analysis. 

In preparing to train to the ANI-1x dataset, a hyperparameter search was performed.. First, models were trained on the QM9 dataset with features which ranged from 1-3 interaction layers, 10, 30, 40, 60, 120 features and 1-3 on-site layers. Additionally, both HIP-NN and HIP-NN-TS models were trained for comparison. QM9 was chosen due to its smaller size, which reduced computation cost. The best models of this set were determined based on MAE test error. These models were retrained, in order to improve error, with a different dataset split. 105426 molecules were used for training, 5000 for validation, and 20397 for testing.  Following this, a similar hyperparameter search was run for ANI-1ccx, however the search was modified using the results from the QM9 search (1-2 interactions, 10, 20, 30, 40, 60, 120 features, 2-4 on-site layers, split 80\% train, 10\% validation, 10\% test). The best results were then taken for 1 and 2 interactions to train the models for the ANI-1x dataset. For QM7-X, the data split uses unknown molecules given by Ref.~\cite{spookynet}, a random holdout of 10,100 conformations for testing, a random 101,000 conformations for validation, and the rest for training. For Allegro-Ag, 80\% of configurations were used for training, 10\% for validation, and 10\% for testing. For ANI-1x and QM7-X, hyperparameter combination used was 2 interaction, 128 features, and 4 on-site layers. A batch size of 512 was used. Allegro-Ag models used $n_\mathrm{int}=1$ and $n_\mathrm{feature}=8$, and a batch size of 4.The number of sensitivities $n_\nu = 20$ was used for QM7, QM9, ANI-1ccx, and ANI1-x, with $n_\nu=30$ on QM7-X, and $n_\nu=16$ for Allegro-Ag.  all trained models. Distance parameters for the sensitivity functions are characterized by three numbers: the soft \emph{lower} cutoff of the innermost sensitivity function, the soft \emph{upper} cutoff of the outermost sensitivity function, and the \emph{hard} cutoff after which all interactions are suppressed using a cosine-squared cutoff function. Distance cutoffs were modified for different datasets. For QM7, the cutoffs were 1.67 Bohr (lower cutoff), 10 Bohr (upper cutoff), and 12.5 Bohr (hard upper cutoff). For QM9 and ANI-1ccx, the cutoff parameters were 0.85 \angstrom (lower), 5 \angstrom (upper), and 5.5 \angstrom (hard). For ANI-1x, cutoffs of 0.75 \angstrom (lower), 5.5 \angstrom (upper), and 6.5 \angstrom (hard) were used. For QM7-X, cutoffs of 0.6 \angstrom (lower), 5.5 \angstrom (upper), and 6.5 \angstrom (hard) were used. For Allegro-Ag, cutoffs of 2.3 \angstrom (lower), 3.75 \angstrom (upper), and 4.0 \angstrom (hard) were used. 

Appendix Table~\ref{tab:allegro_ag_performance} Shows performance of HIP-NN and HIP-NN-TS models in comparison to the original work of Allegro\cite{allegro}. For molecular dynamics, we followed the methods of Ref.~\onlinecite{allegro}. Each frame of molecular dynamics contains more than ten times the number of force evaluations found in the training dataset. In large-scale systems over long enough time-scales, occasionally a pair of atoms drift closer together than any pair found in the training dataset (approximately $2.3\angstrom$). To correct for this, we added pair-wise repulsion using the ZBL potential with an upper cutoff of $2.3\angstrom$.

\begin{table*}
\centering
\begin{tabular}{l|c|c|c|}
\toprule
Metric & $E$ ($\frac{\mathrm{meV}}{\mathrm{atom}}$) & $F$ ($\frac{\mathrm{meV}}{\angstrom}$) & Parameters  \\
\midrule
HIP-NN &  0.466 & 20.9 & 746 \\
HIP-NN-TS, $\ell=1$ &  0.390 & 19.0 & 754\\
HIP-NN-TS, $\ell=2$ & 0.385 & 16.8 & 762\\
Allegro\cite{allegro} &  0.397 & 16.8 & 9K \\

\bottomrule
\end{tabular}
\caption{Mean Absolute Errors (MAE) of HIP-NN, HIP-NN-TS, and Allegro models trained to the Allegro-Ag dataset for both energy $E$ and forces $F$. HIP-NN performs somewhat worse than Allegro, where HIP-NN-TS models provide competitive performance at $\ell=1$ and slightly better performance at $\ell=2$. }
    \label{tab:allegro_ag_performance}
\end{table*}

\begin{figure*}
    \centering
    \includegraphics[width=0.9\textwidth]{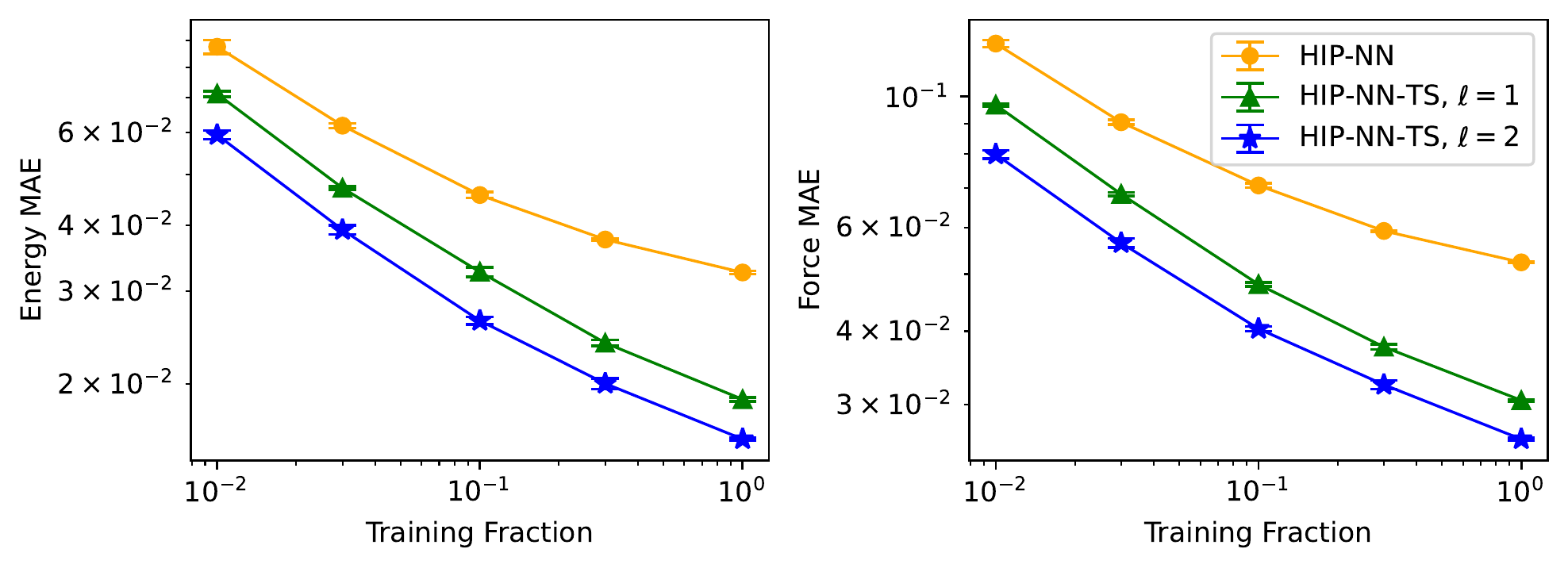}
    \caption{Learning curves for QM7-X showing predictions on the test fold. Energy Errors are given in terms of meV, and Force Errors are given in terms of $\frac{\mathrm{meV}}{\angstrom.}$ The fraction of the data is relative to the maximum training size of 4,074,037 configurations, with other configurations held out for validation, test, and unknown molecules. Error bars show the standard error of the mean over several random seeds.}
    \label{fig:qm7x_learningcurve}
\end{figure*}

The learning curve for QM7-X, showing the model performance of HIP-NN and HIP-NN-TS models as a function of training fraction, is shown in Figure.~\ref{fig:qm7x_learningcurve}

\section{Additional information on ANI1-x/comp6 study}

To assess energy predictions, we measure the MAE and RMSE accuracy for predicting the total energy $E$, and for predicting the energy difference $\Delta E$ between a given conformation and the relaxed geometry.  Total energy errors $E$ are generally larger than conformational energy errors $\Delta E$, and this is particularly noticeable for the ani\_md subset, which contains the largest test systems. This indicates the presence of systematic model errors, which become more noticeable in larger systems. This effect has been witnessed in other datasets using other network architectures\cite{spookynet}. It could be a sign that the model would not perform as well for chemical reactions or isomerizations which move a system between very different basins in the potential energy surface, but more testing would be required to confirm this. In any case, the DFT functional used for the ANI-1x and COMP6 data does not well-represent these types of energies~\cite{Smith2019ApproachingLearning}, so it would not be appropriate to apply these models to the context of reactions. 

\begin{table*}
\centering
\begin{tabular}{l|r|r|r|r|r|r|}
\toprule
Error Type &            $E$ MAE &            $E$ RMSE &     $\Delta E$ MAE &    $\Delta E$ RMSE &            $F$ MAE &           $F$ RMSE \\
\midrule
all        &  $1.261 \pm 0.026$ &   $2.723 \pm 0.226$ &  $1.105 \pm 0.006$ &  $1.951 \pm 0.053$ &  $1.436 \pm 0.008$ &  $2.413 \pm 0.012$ \\
ani\_md    &  $6.478 \pm 0.913$ &  $14.644 \pm 2.417$ &  $2.051 \pm 0.064$ &  $3.863 \pm 0.245$ &  $1.577 \pm 0.027$ &  $2.595 \pm 0.068$ \\
drugbank   &  $1.873 \pm 0.089$ &   $2.808 \pm 0.110$ &  $1.209 \pm 0.004$ &  $1.703 \pm 0.008$ &  $1.356 \pm 0.006$ &  $2.157 \pm 0.010$ \\
gdb10-13   &  $1.420 \pm 0.017$ &   $2.008 \pm 0.015$ &  $1.397 \pm 0.009$ &  $2.027 \pm 0.011$ &  $1.749 \pm 0.010$ &  $2.755 \pm 0.015$ \\
gdb7-9     &  $0.552 \pm 0.014$ &   $0.763 \pm 0.015$ &  $0.540 \pm 0.005$ &  $0.796 \pm 0.008$ &  $0.974 \pm 0.008$ &  $1.509 \pm 0.012$ \\
s66x8      &  $0.976 \pm 0.053$ &   $1.409 \pm 0.075$ &  $0.743 \pm 0.038$ &  $1.142 \pm 0.069$ &  $0.618 \pm 0.006$ &  $1.096 \pm 0.012$ \\
tripeptide &  $1.529 \pm 0.124$ &   $2.137 \pm 0.121$ &  $0.822 \pm 0.016$ &  $1.690 \pm 0.175$ &  $1.055 \pm 0.007$ &  $3.593 \pm 0.119$ \\
\bottomrule
\end{tabular}
    \caption{Single-model errors of HIP-NN-TS with $n_\mathrm{int}=2$ and $\ell_\mathrm{max}=2$ trained to ANI-1x, and tested on COMP6. Energies are reported in $\kcpm$, and forces in $\kcpmpa$. The center value is calculated as the mean over the 8 single-model errors. Error bars are determined by the standard error of the mean over 8 models.}
    \label{tab:comp6_singlemodel_errors}
\end{table*}

\begin{table*}
\centering
\begin{tabular}{l|r|r|r|r|r|r|}
\toprule
Error Type & $E$ MAE & $E$ RMSE & $\Delta E$ MAE & $\Delta E$ RMSE & $F$ MAE & $F$ RMSE \\
Subset     &         &          &                &                 &         &          \\
\midrule
all        &   1.059 &    2.438 &          0.927 &           1.584 &   1.124 &    1.917 \\
ani\_md    &   6.249 &   14.387 &          1.614 &           2.911 &   1.105 &    1.763 \\
drugbank   &   1.534 &    2.297 &          1.002 &           1.409 &   1.052 &    1.679 \\
gdb10-13   &   1.192 &    1.715 &          1.200 &           1.752 &   1.394 &    2.215 \\
gdb7-9     &   0.440 &    0.628 &          0.459 &           0.688 &   0.755 &    1.191 \\
s66x8      &   0.802 &    1.151 &          0.505 &           0.785 &   0.382 &    0.700 \\
tripeptide &   1.271 &    1.726 &          0.661 &           1.388 &   0.812 &    3.049 \\
\bottomrule
\end{tabular}
\caption{Errors of HIP-NN-TS ensemble with $n_\mathrm{int}=2$, $\ell_\mathrm{max}=2$  trained to the ANI-1x dataset, and applied to the COMP test set. Energies $E$ and conformational energies $\Delta E$ are reported in $\kcpm$, and forces $F$ in $\kcpmpa$.}
    \label{tab:comp6_ensemble}
\end{table*}

\begin{figure*}
    \centering
    \includegraphics[width=0.9\textwidth]{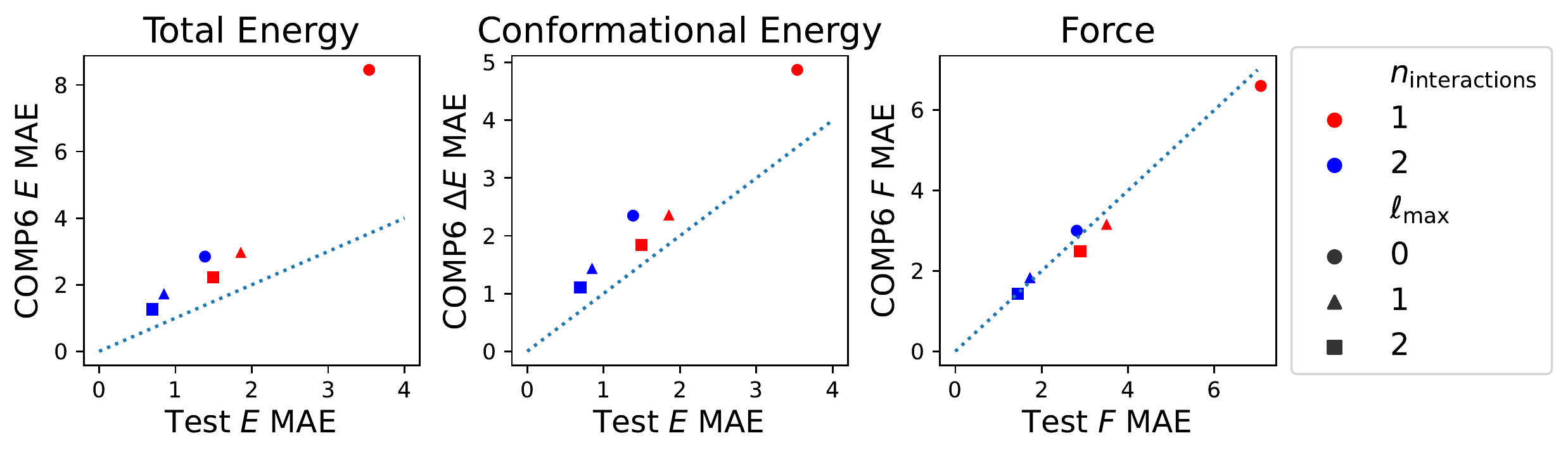}
    \caption{ANI-1x test set errors  vs. COMP6 extensibility set errors, for energy $E$, conformational energy $\Delta E$, and force $F$. These plots demonstrate that improvements on the tests of held-out ANI1-x data yielded similar percentage-wise improvements on COMP6 tests. Each point is determined by averaging the performance metrics over all neural networks in the ensemble.}
    \label{fig:extensibility_vs_test}
\end{figure*}

Table \ref{tab:comp6_singlemodel_errors} indicates the performance of single models on subsets of the COMP6 benchmark suite.

Figure \ref{fig:extensibility_vs_test} shows the correlation between test set error (ANI-1x, held out conformations) and extensibility error (COMP6 benchmark) for energy, conformational energy, and force predictions.

\begin{figure*}
    \centering
    \includegraphics[width=1.0\textwidth]{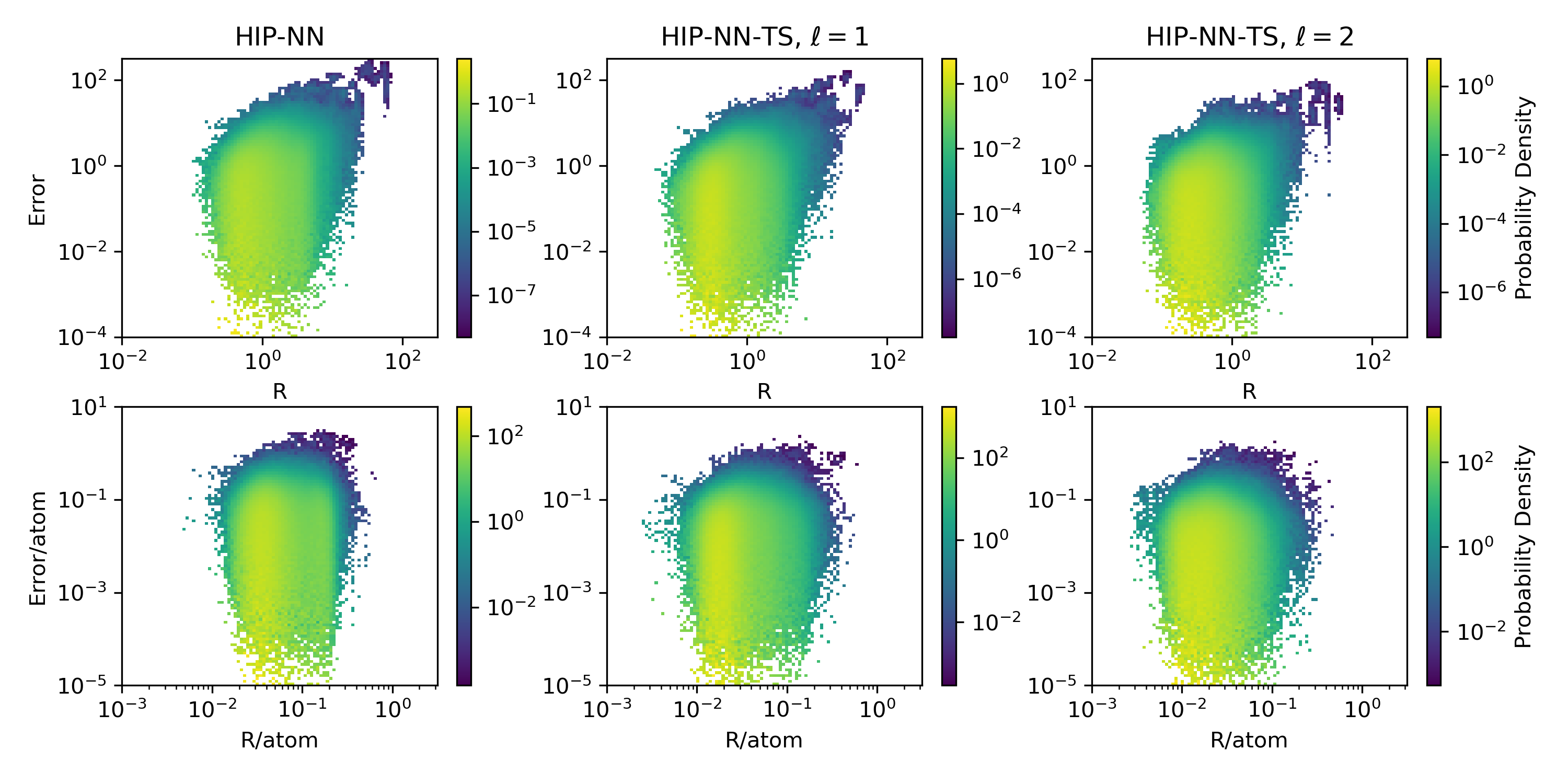}
    \caption{Hierarchical parameter $R$ (non-dimensional) compared to absolute energy error ($\kcpm$), represented as a 2D histogram, on the predictions of the COMP6 test set. The top row of plots is extensive, and the bottom row of plots compares the variables on a per-atom basis. Each column shows models trained with a different tensor order. Each plot shows the combined predictions of all eight models in each ensemble.}
    \label{fig:hierarchicality}
\end{figure*}

\section{Alternative Forms for Tensor Sensitivity}
\label{sec:alternative_ts}

A simple form of regularizing Eq.~\ref{eq:hip-vec-1} is to replace the absolute magnitude with a regularized magnitude, $K(\boldsymbol{x},\epsilon) = \sqrt{||\boldsymbol{x}||^2+\epsilon^2}$, so that the interaction equation becomes

\begin{equation}
\mathcal{I}_{i,a}^{\textrm{HIP-NN-TS-$\epsilon$}}=\boldsymbol{\mathcal{E}}^{(0)}_{i,a}+t^{(1)}_a K(\boldsymbol{\mathcal{E}}_{i,a}^{(1)},\epsilon)+t^{(2)}_a K(\boldsymbol{\mathcal{E}}_{i,a}^{(2)},\epsilon )+\dots\label{eq:hip-vec-1-epsilon}
\end{equation}
The present work has utilized this version with numerically small $\epsilon=10^{-15}$ to avoid undefined second-derivatives associated with training to forces. To avoid the cusp entirely, one might choose larger $\epsilon$ which more strongly smooths out the cusp.

In future work, one might also consider replacing Eq.~(\ref{eq:hip-vec-1}) with the alternative definition,
\begin{equation}
\mathcal{I}_{i,a}^{\textrm{HIP-NN-TS-ALT}}=\sum_{\ell=0}^{\ell_\mathrm{max}} t_a^{(\ell)}|\boldsymbol{\mathcal{E}}_{i,a}^{(\ell)}|^{2},\label{eq:hipalt2}
\end{equation}
which would treat the scalar case $\ell = 0$ symmetrically with higher tensor orders $\ell > 0$, and would avoid introduction of cusp singularities. An advantage is that the potential energy surface would become uniformly smooth. However, there are also disadvantages. First, the model capacity would not be a strict superset of scalar HIP-NN. Second, the scalar response would no longer distinguish positive and negative messages. Third, the model would be quadratic rather than linear in large input feature vectors $z$ and weight values $v$. This final point merits more discussion. The authors of Cormorant noted that quadratic nonlinearities may be a source of instability in their training~\cite{Anderson2019Cormorant:Networks}; quadratic nonlinearities have constant curvature and so gradients systematically grow with the values of the activations. It is quite possible that this instability is an instance of the exploding gradient problem, wherein gradient values during training systematically grow with network depth, preventing smooth optimization. In the neural network literature, it has been observed that linear response for large input values can be beneficial~\cite{pmlr-v9-glorot10a}. This was a key motivation behind the resurgence of rectifying activation functions, such as the Rectified Linear Unit (ReLU) and the Softplus function. These activation functions yield more uniform gradient flow during training~\cite{pmlr-v15-glorot11a,pmlr-v9-glorot10a} compared to earlier activation functions which saturate and therefore suffer from the vanishing gradient problem. In light of these facts, HIP-NN-TS uses linear tensor norms, so that individual layers respond approximately linearly to large input values. No training instability was observed in HIP-NN-TS.

A more targeted study would be necessary to determine the impacts of various possible modifications to the HIP-NN-TS activation formula, Eq.~\eqref{eq:hip-vec-1}. An interesting fact to consider is that non-smooth potential energies could arise physically from the crossing of degenerate energy levels; it may be that the conical cusp in HIP-NN-TS has value for describing this type of phenomenon. It has recently been pointed out that square-root nonlinearities have a long history in the development of interatomic potentials, and have demonstrated benefits to the performance of the ACE model~\cite{AtomicClusterExpansion}.

\bibliographystyle{apsrev4-1}

\end{document}